\documentclass{aa}

\usepackage{acronym}  
\usepackage{amssymb} 
\usepackage{amsbsy}
\usepackage{amsmath}
\usepackage{longtable}
\usepackage{multirow}
\usepackage{units}
\usepackage{bbold}
\usepackage{xcolor}
\usepackage{natbib}

\defcitealias{Graber2024}{Paper I}
\acrodef{PMPS}[PMPS]{Parkes Multibeam Pulsar Survey}
\acrodef{SMPS}[SMPS]{Swinburne Intermediate-latitude Pulsar Survey}
\acrodef{HTRU}[HTRU]{High Time Resolution Universe}
\acrodef{ISM}[ISM]{interstellar medium}
\acrodef{CNN}[CNN]{convolutional neural network}
\acrodef{MDN}[MDN]{mixture density network}
\acrodef{GMM}[GMM]{Gaussian-mixture model}
\acrodef{ReLU}[ReLU]{rectified linear unit}
\acrodef{SBI}[SBI]{}
\acrodef{MCMC}[MCMC]{Markov chain Monte Carlo}
\acrodef{DM}[DM]{dispersion measure}
\acrodef{SNR}[SNR]{signal-to-noise ratio}
\acrodef{NPE}[NPE]{Neural posterior estimation }
\acrodef{NLE}[NLE]{neural likelihood estimation}
\acrodef{NRE}[NRE]{neural ratio estimation}
\acrodef{CI}[CI]{credible interval}
\acrodef{HDR}[HDR]{highest-density region}
\acrodef{KDE}[KDE]{kernel-density estimation}
\acrodef{ISM}[ISM]{interstellar medium}
\acrodef{TSNPE}[TSNPE]{truncated neural sequential posterior estimator}

\newcommand{\bt}{\boldsymbol{\theta}}

\newcommand{\bx}{\boldsymbol{x}}
\newcommand{\pr}{\mathcal{P}}

\begin{document}

   \title{Radio pulsar population synthesis with consistent flux measurements using simulation-based inference }
    \titlerunning{Radio pulsar population synthesis with consistent flux measurements using SBI} 

   \author{Celsa Pardo-Araujo
          \inst{1,2}
          \and
          Michele Ronchi\inst{1,2}
          \and Vanessa Graber \inst{3}
          \and Nanda Rea \inst{1,2}
          }

   \institute{Institute of Space Sciences (CSIC-ICE), Campus UAB, Carrer de Can Magrans s/n, 08193, Barcelona, Spain
         \and
             Institut d'Estudis Espacials de Catalunya (IEEC), Carrer Gran Capit\`a 2--4, 08034 Barcelona, Spain
        \and
            Department of Physics, Royal Holloway, University of London, Egham, TW20 0EX, UK
             }

   \date{}

 
  \abstract
   {The properties of isolated Galactic radio pulsars can be inferred by modelling their evolution, from birth to the present, through pulsar population synthesis. This involves simulating a mock population, applying observational filters, and comparing the resulting sources to the limited subset of detected pulsars. We specifically focus on the magneto-rotational properties of Galactic isolated neutron stars and provide new insights into the intrinsic radio luminosity law. To better constrain the intrinsic radio luminosity, for the first time in pulsar population synthesis studies, we incorporate data from the Thousand Pulsar Array (TPA) program on MeerKAT, which contains the largest unified sample of neutron stars with consistent flux measurement to date. In particular, we employed a simulation-based inference (SBI) technique called Truncated sequential neural posterior estimation (TSNPE) to infer the parameters of our pulsar population model. This technique trains a neural density estimator on simulated pulsar populations to approximate the posterior distribution of underlying parameters. This method efficiently explores the parameter space by focusing on regions most likely to match the observed data, significantly reducing the required training dataset size. We find that adding flux information as an input to the neural network significantly improves the constraints on the pulsars' radio luminosity and improves the estimates on other input parameters. Moreover, we demonstrate the efficiency of TSNPE over standard neural posterior estimation (NPE), as we achieve robust inferences of magneto-rotational parameters consistent with previous studies while using only around $4\%$ of the simulations required by NPE approaches.}


   \maketitle
%

\section{Introduction}
\label{sec:intro}

Neutron stars are remnants of core-collapse supernovae of massive stars characterised by their extreme properties, such as their high magnetic field and rapid rotation. Originally detected in radio as pulsars, neutron stars have since been observed across the electromagnetic spectrum, from radio to gamma rays. This diverse set of observations has resulted in a wide variety of classes of neutron stars, whose many properties, such as magnetic field evolution, are still not fully understood. Among the various classes of neutron stars, radio pulsars are the most prevalent. These isolated neutron stars emit primarily in the radio band. As they rotate, their strong magnetic fields accelerate particles to relativistic speeds along open magnetic field lines, producing the characteristic beamed emission observed in pulsars \citep{Goldreich1969,Ruderman1975}. However, the precise mechanisms driving this coherent radio emission remain poorly understood \citep{Zhang2000, Chen1993}.  As a result, studying neutron stars not only deepens our understanding of their nature but also provides valuable insights into broader astrophysical phenomena, including core-collapse supernovae.\\
Pulsar population synthesis studies are a powerful tool to analyse the whole population of neutron stars. Although we expect between $10^{6}$ and $10^{8}$ neutron stars in our galaxy based on core-collapse supernova rates \citep{Rozwadowska2021}, only approximately $3,000$ isolated neutron stars have been detected \citep{Manchester2005} due to observational biases. By modelling neutron stars from their birth to the present and applying observational filters, population synthesis allows us to compare the observed neutron stars with simulated populations and thus to study the properties of the whole population by looking at the small subset of detected sources. Numerous studies have applied this technique to model the Galactic neutron star population using various statistical techniques \citep[e.g.,][]{Narayan1990, Lorimer2004, Faucher2006, Gonthier2007, Bates2014, Gullon2014, Gullon2015, Cieslar2020}. However, these studies typically rely on simplified likelihoods or metrics to compare simulations with observations, followed by methods such \ac{MCMC} as applied in \citep{Cieslar2020} or annealing techniques \citep{Gullon2014} for parameter estimation.\\
In \cite{Graber2024} (hereafter referred to as \citetalias{Graber2024}), we demonstrate that simulation-based inference \citep[SBI; see, e.g.,][]{Cranmer2020} is an effective technique for performing inference with complex simulators, such as those modelling the evolution of the neutron star population. \acs{SBI}, also known as likelihood-free inference, uses neural networks and the stochastic nature of simulators to approximate the posterior distribution without the need for a simplified likelihood model. \citetalias{Graber2024} shows the strength of this method when performing pulsar population synthesis with a realistic prescription for magnetic field decay, allowing us to infer the magnetic field and period distributions at birth.\\
With the future Square Kilometre Array (SKA), the number of detected pulsars in our Galaxy will increase by an order of magnitude \citep{Dai2017}, which will be crucial for deepening our understanding of the neutron star population. In particular, MeerKAT has already created the largest unified sample of neutron stars to date, including consistent flux measurements \citep{Posselt2023}. Building on the work in \citetalias{Graber2024}, we expand the methodology in the present paper by taking advantage of the new accurate flux measurements. Using the pulsar population synthesis code described below, we constrain the magnetic field and period at birth and also gain new insights into the radio luminosity law of isolated Galactic neutron stars.\\
In \citetalias{Graber2024}, we employed an SBI method known as Neural posterior estimation \citep[NPE;][]{Papamakarios2016}, which uses a neural network to directly approximate the posterior distribution of the underlying parameters. However, \acs{NPE} becomes inefficient when dealing with high-dimensional parameter spaces. In this work, we address this limitation by using a more efficient algorithm called Truncated sequential neural posterior estimation \citep[TSNPE;][]{Deistler2022}. In particular, \acs{TSNPE} guides the exploration of the parameter space by using the neural network itself, making the inference process significantly more efficient. This efficiency allows us to expand the number of parameters being inferred and extend the scope of our analysis.\\
The paper is organized as follows: Section 2 summarises the most relevant aspects of our population synthesis framework used in this study. We then
provide an overview of SBI and \acs{TSNPE} in Sections 3.1 and 3.2, respectively, whereas in Sections 3.3 to 3.5, we summarise the technical aspects of TSNPE. In Section \ref{subsec:exps}, we explain the experiments performed in this work. We show the results for the experiments used to test the TSNPE technique in Sections \ref{subsec:test-5} and \ref{subsec:test-7}, while we outline the main results in Section \ref{subsec:main_results}. Finally, we provide a detailed discussion of our approach and results in Section \ref{sec:conclusions}.

\section{Pulsar population synthesis}
\label{sec:popsyn}

In this work, we focus on modelling the population of isolated rotation-powered radio pulsars. These are isolated neutron stars whose radio emissions are powered by the loss of rotational energy. We performed Monte Carlo simulations to model both the dynamical and magneto-rotational evolution of this type of neutron star, which we followed by the application of observational biases to compare the simulated populations with observations. For a detailed description of all equations and the physics underlying the simulations, we refer the reader to \citetalias{Graber2024}, though we summarise the main steps here for completeness.\\
\subsection{Dynamical and magneto-rotational evolution}
To reduce the computational cost, we took advantage of the fact that the magneto-rotational and dynamical evolution are decoupled. This allowed us to fix the parameters governing the dynamical evolution, which in turn enabled us to perform the corresponding evolution only once to generate a comprehensive database of dynamically evolved neutron stars. This database then served as the foundation for the subsequent magneto-rotational evolution.  \\
To create the dynamically evolved database, we simulated a population of $10^7$ neutron stars, assigning them random ages up to $10^8$ yr to ensure a realistic detection sample, as neutron stars older than this age are no longer observable. Since we expected massive OB stars, the precursors of neutron stars, to be located in star-forming regions directly related to the distribution of free electrons, we sampled the initial positions based on the Galactic electron density distribution from \citet{Yao2017}. During supernova explosions, neutron stars receive kicks as a result of asymmetries in the explosion \citep[see][]{Coleman2022, Janka2022}. To model these kicks, we sample the kick velocity from a Maxwell distribution, using a fiducial value of $\sigma_{\rm k} \approx \unit[260]{km \, s^{-1}}$ for the dispersion parameter  \citep{Hobbs2005}, which is broadly consistent with the observed proper motions of radio pulsars \citep{Hobbs2005, Faucher2006}. To determine the positions of neutron stars in the Galaxy at any given time, we then computed their dynamical evolution by solving the Newtonian equations of motion in Galactocentric coordinates.  \\
Radio pulsars can be approximated as rotating magnetic dipoles, with the magnetic dipole axis misaligned with respect to the rotation axis. As neutron stars age, magnetospheric torques cause a gradual loss of rotational energy. This torque leads to a gradual slowdown in the star's spin and drives the alignment of the magnetic and rotation axes. We simulated this magneto-rotational evolution by sampling neutron stars from our dynamical database and tracking changes in spin period, misalignment angle, and magnetic field.\\
We began by randomly selecting an initial misalignment angle, $\chi_0$, in the range $[0, \pi/2]$ based on the probability distribution \citep{Gullon2014}
\begin{equation}
	\mathcal{P}(\chi_0) = \sin \chi_0.
\end{equation}
 Next, we selected the initial magnetic field strength, $B_0$ (in G), and the initial spin period, $P_0$ (in s), for each neutron star. These parameters are drawn from normal distributions in logarithmic space defined as follows \citep{Popov2010, Gullon2014, Igoshev2020, Igoshev2022, Xu2023}:
\begin{align}
	\mathcal{P}(\log B_0) &= \frac{1}{\sqrt{2 \pi} \sigma_{\log B_0}}
		\, \exp\left(-\frac{(\log B_0 - \mu_{\log B_0})^2}{2 \sigma_{\log B_0}^2} \right),
			\label{eqn:B_pdf} \\[1.8ex]
  	\mathcal{P_0}(\log P_0) &= \frac{1}{\sqrt{2 \pi} \sigma_{\log P_0}}
      		\, \exp\left(-\frac{(\log P_0 - \mu_{\log P_0})^2}{2 \sigma_{\log P_0}^2} \right).
			\label{eqn:P_pdf}
\end{align}
The means, $\mu_{\log B}, \mu_{\log P}$, and the standard deviations, $\sigma_{\log B}, \sigma_{\log P}$, are free parameters in our model. \\
To model the evolution of the spin period and the misalignment angle due to the loss of rotational energy via dipole radiation, we followed \citet{Philippov2014} and \citet{Spitkovsky2006}
 and solved the following coupled differential equations:
\begin{align}
	\dot{P} &= \frac{\pi^2}{c^3}\frac{B^2 R_{\rm NS}^6}{I_{\rm NS} P} \left( \kappa_0 + \kappa_1 \sin^2 \chi \right),
		\label{eqn:P_ode} \\[1.8ex]
	\dot{\chi} &= -\frac{\pi^2}{c^3}\frac{B^2 R_{\rm NS}^6}{I_{\rm NS} P^2} \, \kappa_2 \sin\chi \cos\chi,
		\label{eqn:chi_ode}
\end{align}
where $c$ is the speed of light, $R_{\rm NS} \approx \unit[11]{km}$ is the neutron-star radius, and $I_{\rm NS} \simeq 2 M_{\rm NS} R_{\rm NS}^2 / 5 \approx \unit[1.36 \times 10^{45}]{g \, cm^{2}}$ is the stellar moment of inertia (for a fiducial mass $M_{\rm NS} \approx \unit[1.4]{M_{\odot}}$). For realistic pulsars surrounded by plasma-filled magnetospheres, we chose $\kappa_0 \simeq \kappa_1 \simeq \kappa_2 \simeq 1$.\\
We also incorporated magnetic field decay to accurately capture the long-term magneto-rotational evolution of neutron stars. The evolution of the dipolar magnetic field strength in neutron stars is influenced by both the Hall effect and ohmic dissipation in the crust \citep{Vigano2013}. This decay is particularly significant for strongly magnetised neutron stars (fields above $~10^{13}$ G), making it relevant for a considerable fraction of our simulated pulsar population. We compute the dipolar magnetic field at time $t$ by parametrising numerical simulation obtained with the magneto-thermal code of \cite{Vigano2021}. However, at late times ($t> 10^6$ yr) this numerical prescription becomes unreliable because it relies on implementations of complex microphysics that
are unsuitable for cold and old stars. Instead, we modelled the evolution of the magnetic field decay at late times with a power law:
\begin{equation}
	B(t) \propto \left(1+ \frac{t}{\tau_{\rm late}} \right)^{a_{\rm late}},
		\label{eqn:B_late}
\end{equation}
where $\tau_{\rm late} \approx \unit[2 \times 10^6]{yr}$ was chosen to fit the simulations and the power-law index, $a_{\rm late}$, is an additional free parameter of our model (see Appendix A of \citetalias{Graber2024} for details). Therefore, we have a total of five free parameters related to the magneto-rotational evolution, which hereafter is referred to as the magneto-rotational parameters.

\subsection{Intrinsic bolometric luminosity prescription}

To compare the simulated populations with observations, it is essential to consider observational biases. We begin by computing the geometry of the radio beam to estimate how many neutron stars have radio beams that cross our line of sight  \citepalias[see Section 2.4 in][for details]{Graber2024}. 
For those neutron stars that are potentially detectable, we then compute their radio fluxes to determine if they fall within the detection limits of the modelled radio surveys. To obtain these radio fluxes, we assume that some percentage of the rotational energy, $\rm E_{rot}$, of each star is converted into coherent radio emission. Therefore, we modelled the intrinsic luminosity as being proportional to a power law of the rotational energy loss, with an exponent $\alpha$:
\begin{equation}
    L_{\rm int} = L_0 \left( \frac{ \dot{E}_{\rm rot}}{ \dot{E}_{0,\rm rot}} \right)^{\alpha},
    \label{eqn:luminosity}
\end{equation}

\noindent where $L_0$ is a normalisation factor whose logarithm is sampled from a normal distribution with mean $\mu_{\log L}$ and standard deviation $\sigma_{\log L} = 0.8$ \citep[see also][]{Faucher2006, Gullon2014}. $\dot{E}_{0,\rm rot}$ represents the rotational energy loss corresponding to a neutron star luminosity $L_0$. Although this prescription differs slightly from that of \citetalias{Graber2024} (Eq. 16),  both relate the loss of the rotational energy with the luminosity. We use this new prescription in this work to remove correlations between the power law index $\alpha$ and the normalisation factor $L_0$, which helps with the inference procedure outlined below. We choose  $\dot{E}_{0,\rm rot} = \unit[10^{29}]{erg \, s^{-1}}$ to be the minimum observed value for the radio pulsar population \citep[see Figure 3 of][]{Posselt2023}. In this work, $\mu_{\log L_0}$ and $\alpha$ are treated as free parameters, resulting in a total of seven free parameters, including those related to the magneto-rotational evolution.   

\subsection{Radio survey detectability}
In what follows, our inference of the luminosity parameters depends on the details of the emission geometry as well as the propagation of the radiation. We summarise our prescription here and refer the reader to Section 2.4 of \citetalias{Graber2024} for further information. In particular, once we computed the luminosity and the geometry of the beam and extracted the pulsar distances from the dynamical database, we computed the bolometric flux, S, for a given pulsar:
\begin{equation}
	S = \frac{L_{\rm int}}{\Omega_{\rm b} d^2}, 
\end{equation}
where $d$ is the distance and $\Omega_{\rm b}$ the solid angle covered by a pulsar’s two radio
beams. To calculate the radio flux density, $S_{f}$ (in Jy), at a given observing frequency, $f$, from the bolometric flux, S, we followed the approach of \citet{Lorimer2012}, assuming that the radio emission spectrum follows a power law in $f$. We note that we adopt two different spectral indices in the following: for the simulations used in the experiments constraining five model parameters, where we compare the results of this work with those of \citetalias{Graber2024} (see Section \ref{subsec:exps} below), we adopt a spectral index of $-1.6$ as suggested by \citet{Jankowski2018}. In contrast, for experiments focused on inferring magneto-rotational and luminosity parameters by adding MeerKAT fluxes, we assume a spectral index of $-1.8$ based on the mean spectral index estimated by \citet{Posselt2023} (see their Figure 8) to maintain consistency. At this point, given the intrinsic pulse width $w_{\rm int}$, we can compute the fluence as $S_{f} w_{\rm int}$. Although this fluence is conserved when the radio signal propagates towards the observer, the pulse gets broadened by dispersion and scattering caused by the free electrons and the interstellar medium. Therefore, we computed the radio fluxes reaching Earth as:
\begin{equation}
	S_{f , {\rm obs}} \simeq S_{f} \frac{ w_{\rm int}}{w_{\rm obs}},
	\label{eqn:obs_flux} 
\end{equation}
where $w_{\rm obs}$ is the observed pulse width.\\
In this study, we focus on three major surveys conducted with Murriyang, the Parkes radio telescope: the \acf{PMPS} \citep{Manchester2001, Lorimer2006}, the \acf{SMPS} \citep{Edwards2001, Jacoby2009}, and the low- and mid-latitude \acf{HTRU} surveys \citep{Keith2010}. 

Equipped with the observed radio fluxes at the central frequency of each radio survey, we can now determine whether each pulsar in our synthetic sample is detectable by each radio survey. First, we exclude stars that do not lie within the survey's sky coverage. Next, we computed the signal-to-noise ratio using the radiometer equation \citep{Lorimer2012}:
\begin{equation}
        S/N = \frac{ S_{\rm mean} G \sqrt{n_{\rm pol} \Delta f_{\rm bw} t_{\rm obs}} }
        		{ \beta \left[ T_{\rm sys} + T_{\rm sky }(l, b) \right] } 
		\sqrt{\frac{P- w_{\rm obs}}{w_{\rm obs}}}.
		\label{eqn:radiometer}
\end{equation}
Here, $S_{\rm mean} \simeq S_{f, {\rm obs}} w_{\rm obs} / P$ denotes the mean flux density averaged over a single rotation period $P$, $G$ is the receiver gain \citep[see][for details]{Lorimer1993, Bates2014}, $n_{\rm pol}$ is the number of detected polarisations, $\Delta f_{\rm bw}$ the observing bandwidth, $t_{\rm obs}$ the integration time and $\beta > 1 $ a degradation factor that accounts for imperfections during the digitisation of the signal. Moreover, $T_{\rm sys}$ denotes the system temperature and $T_{\rm sky}(l, b)$ is the sky background temperature dominated by synchrotron emission of Galactic electrons. Therefore, we count a pulsar as detected if its computed S/N ratio exceeds the detection threshold for that survey.
A summary of all relevant survey parameters is provided in Table 1 in \citetalias{Graber2024}.\\
To efficiently perform the magneto-rotational evolution without assuming a fixed birth rate, we adopt the following approach: we sample a batch of $100,000$ neutron stars from our dynamical database, evolve this subset magneto-rotationally, and subsequently compute their radio emission properties. We then evaluate how many of these neutron stars are detected by each survey and check if this number matches the observed population. This process is repeated until the number of detected pulsars in the simulation matches that of the observed sample for each survey. As the number of simulated detected neutron stars approaches that of the observations, we reduce the batch size to closely match the observed number of neutron stars. We note that this method allows us to estimate the birth rate at the end of each simulation by dividing the total number of neutron stars sampled from the dynamical database by the maximum age considered. The resulting output of each simulation consists of three data frames, each containing the detected neutron stars for one of the surveys.

\subsection{Observed neutron star population}
For the observed neutron star sample, we use the ATNF Pulsar Catalogue \citep{Manchester2005}\footnote{\url{https://www.atnf.csiro.au/research/pulsar/psrcat/}}, excluding extragalactic sources and those located in globular clusters. Our focus is on isolated Galactic radio pulsars, as our current model does not account for the physics involved in simulating recycled millisecond pulsars, which have undergone accretion from a companion star. Consequently, we exclude pulsars with period derivatives below $\dot{P} < \unit[10^{-19}]{s \, s^{-1}}$, and $P < \unit[0.01]{s }$ to remove recycled neutron stars.

\begin{figure*}

\centering
\includegraphics[width = 0.6\textwidth]{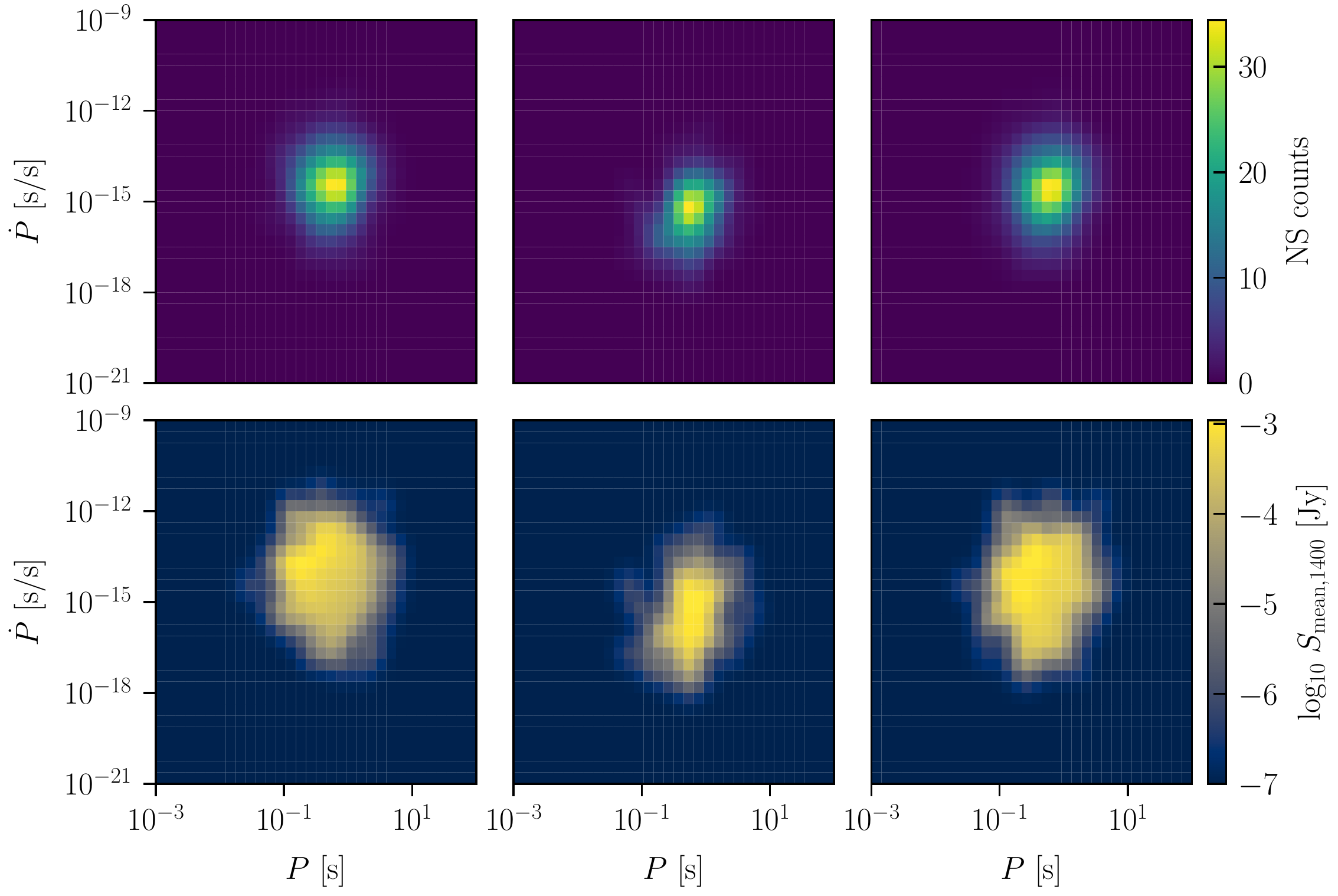}
\caption{Example of the six density maps for a random simulated neutron star population, which we fed into the \acs{SBI} pipeline. The top and the bottom row show the $P$-$\dot{P}$ diagrams and the $P$-$\dot{P}$ averaged flux maps for each of the three surveys, respectively. In the top row, the colour represents the density in neutron star number within each bin, while in the bottom row the colour represents the averaged flux in Jy within each bin. For bins without any stars in the bottom row, the average flux has been set to -7.}
\label{fig:density_maps_test}
\end{figure*}

For Experiments 1 and 2 (for more details see Section \ref{subsec:exps}), we employ version v1.69 of the ATNF Pulsar Catalogue as we compare our results with those from \citetalias{Graber2024} and use their training dataset. However, for Experiments 3 through 5, where we also vary the luminosity and incorporate MeerKAT flux measurements, we use the most recent version of the catalogue v2.5.1. In this updated version, the number of detected pulsars per survey, which we aim to recover in our simulations, is as follows: \begin{align} &\text{\ac{PMPS}: $1045$ observed pulsars}, \nonumber \\
&\text{\ac{SMPS}: $218$ observed pulsars}, \label{eqn:number_pulsars} \\
&\text{\ac{HTRU}: $1037$ observed pulsars}. \nonumber \end{align}

We note that even though we apply the same selection criteria to filter the observed data, the numbers of pulsars differ slightly from those reported in \citetalias{Graber2024}. The discrepancy in the number of detected sources in \ac{HTRU} between this paper and Paper I is due to a bug in version v1.69 of the ATNF Pulsar Catalogue, while the difference in \ac{PMPS} is due to the reprocessing of archival data from \ac{PMPS} \citep{Sengar2023}. Additionally, we have updated the period values for 11 pulsars for which the ATNF Pulsar Catalogue initially reported a harmonic of the true period, following the corrections made by \cite{Song2023}.
For flux measurements, we use data from the Thousand Pulsar Array (TPA) program \citep{Johnston2020a}, which is part of the large survey project MeerTIME on the MeerKAT telescope. The TPA provides a consistently observed sample, as the data were collected using a single telescope with similar observing parameters and processed through a standardised data reduction pipeline. Specifically, we use the flux measurements at $\unit[1.429]{GHz}$  reported by \citet{Posselt2023}. Since the TPA program did not observe all pulsars detected with \ac{PMPS}, \ac{SMPS}, and \ac{HTRU}, we compute the number of overlapping pulsars between MeerKAT and each of these surveys as follows:
 \begin{align}
 &\text{\ac{PMPS} $\&$ MeerKAT: $640$ observed pulsars}, \nonumber \\
 &\text{\ac{SMPS} $\&$ MeerKAT: $170$ observed pulsars}, \label{eqn:numbers_meerkat} \\
&\text{\ac{HTRU} $\&$ MeerKAT: $668$ observed pulsars}. \nonumber
\end{align}

We verified that these subsets of overlapping pulsars do not show additional biases in terms of dispersion measure (DM), sky position, period, or period derivative compared to the full observed populations in the three surveys. Because the TPA program was designed to re-observe known pulsars rather than detect new ones, and TPA does not introduce any additional biases, we do not need to model a separate detection filter for the MeerKAT flux measurements. Instead, we extract a random subset of simulated neutron stars for each survey to match the number of pulsars observed by MeerKAT as listed in \eqref{eqn:numbers_meerkat}, when comparing the simulated fluxes measurements with the observed values.

\subsection{Representation of simulation output}
\label{subsec:density_maps}
To feed the simulated neutron star population into the neural-network based \acs{SBI} algorithm discussed in Section \ref{sec:sbi}, we need to choose a suitable representation. For this purpose, we represent each mock simulation as six density maps: three $P$-$\dot{P}$ diagrams and three $P$-$\dot{P}$ averaged flux maps, corresponding to the three radio surveys modelled in this study. An example of these maps is shown in Figure ~\ref{fig:density_maps_test}. Both types of maps are $P$-$\dot{P}$ diagrams with parameter limits set to $P \in [0.001, 100] \, {\rm s}$ and $\dot{P} \in [10^{-21}, 10^{-9}] \, {\rm s \, s^{-1}}$, using a resolution of 32 bins. However, they differ in terms of colour representation and the number of simulated neutron stars present. In the $P$-$\dot{P}$ density maps, shown in the top panels, the colour indicates the number of neutron stars in each bin, and the number of simulated neutron stars matches the total number of detected pulsars in each survey, as listed in \eqref{eqn:number_pulsars}. On the other hand, the $P$-$\dot{P}$ averaged flux maps, shown in the bottom panels, represent the average flux within each bin matching the numbers listed in \eqref{eqn:numbers_meerkat}. To prevent abrupt changes in pixel intensity in these binned distributions, we have applied a Gaussian smoothing filter with a radius of $4\sigma$ and $\sigma = 1$ \citep[see Chapter 4 of][for more details on Gaussian smoothing filters]{Russ2011}, which also improves stability during the training of our machine-learning pipeline. Throughout this work, the ground truth or labels refer to the parameter value $\bt$ used to generate each simulated population. To further ensure consistency, we standardised the data so that the neural network receives inputs and labels with similar magnitudes. The parameter labels were standardised over the entire parameter range, while each density map was standardised individually. This process ensures that the mean values of each map are centred at 0 with a standard deviation of 1.\\

\section{Simulation-based inference}
\label{sec:sbi}
\subsection{Overview}
Recently, \acs{SBI} has emerged as a powerful tool to perform parameter estimation for complex simulators. Specifically, given a model, parameter estimation consists of computing the free parameters that best reproduce the observed data. As we are interested in computing the uncertainties associated with the estimated parameters, we compute the probability of the parameters, $\bt$, given the data, $\bx$, known as the posterior distribution, $\pr(\bt | \bx)$. The posterior distribution is obtained using Bayes' theorem \citep{Stuart1994} as follows:

\begin{equation}\label{eqn:Bayes_theorem}
	\pr(\bt | \bx) = \frac{\pr(\bt) \pr(\bx |\bt)}{\pr(\bx)},
\end{equation}

\noindent where $\pr(\bt)$ is the prior, representing our initial knowledge of the model parameters. The term $\pr(\bx |\bt)$ is the likelihood, which describes the probability of observing the data given specific parameter values. The denominator, $\pr(\bx)$, known as the evidence, is computed by integrating the likelihood over all possible parameter values:
\begin{equation}
	\pr(\bx) \equiv \int \pr(\bx |\bt') \pr(\bt') \, {\rm d} \bt' .
		\label{eqn:evidence}
\end{equation}

While computing the analytical expression of the posterior distribution is often unfeasible, obtaining samples from the posterior remains possible in many cases. This can be achieved with \ac{MCMC} methods, provided that the likelihood function is known or can be easily approximated. However, for a complex simulator such as ours (see Section \ref{sec:popsyn} for more details), the likelihood is intractable as it is implicitly defined within the simulator. In this sense, the simulator acts as a stochastic generator that, given parameters $\bt$, outputs a mock observation $\bx$. Due to the stochastic nature of the simulator, the output $\bx$ will vary across different runs, even for a fixed set of parameters $\bt$. Therefore, we can use the simulator to generate samples from the likelihood $\pr(\bx | \bt)$ when approximating the posterior distribution.\\
Three approaches are mainly used in \acs{SBI} when employing neural networks: 
\begin{itemize}
    \item Neural posterior estimation: The neural network directly approximates the map between the free parameters of the model, $\bt$, and the posterior distribution, $\pr(\bt | \bx)$ \citep[e.g.][]{Papamakarios2016, Lueckmann2017, Greenberg2019, Mishra-Sharma2022, Vasist2023, Dax2021, Barret2024}. 
    \item Neural likelihood estimation: A neural network is trained to emulate the simulator, i.e., used as a fast likelihood sample generator. Once the network has been trained, it can be used together with a sampling method, such as MCMC, to sample the posterior distribution \citep[e.g.][]{Papamakarios2018, Alsing2019}.
    \item Neural ratio estimation: In this approach, a deep learning classifier learns the likelihood-to-evidence ratio, denoted by $r(\bt,\bx) \equiv \pr(\bx |\bt) / \pr (\bx)$, which is equivalent to $\pr(\bt  |\bx) / \pr (\bt)$ using Bayes' theorem~\eqref{eqn:Bayes_theorem}. To sample from the posterior distribution, one can then use \ac{MCMC} or other sampling algorithms \citep[e.g.][]{Hermans2019, Miller2021, Bhardwaj2023, Saxena2024}.
    
\end{itemize}
In this work, we use \acs{NPE} as we are interested in directly approximating the posterior distribution of the underlying parameters. Generally, \acs{NPE} involves sampling from the prior distribution, $\bt \sim \pr(\bt)$, generating synthetic data, $\bx \sim \pr(\bx|\bt)$, and which we subsequently used to train a neural density estimator. The neural density estimator, $q$, is parametrised by a neural network $F$ with weights $\phi$ (i.e., $q_{F(\bx, \boldsymbol{\phi})}$). The network is optimised by minimising the following loss function
\begin{equation}
	\mathcal{L}(\boldsymbol{\phi}) = - \sum_{i=1}^N \log q_{F(\bx_i, \boldsymbol{\phi})} (\bt_i)
		\label{eqn:loss}
\end{equation}
over a training data set $\{\bt_i, \bx_i\}$ of size $N$. This loss is minimised when the neural density estimator approximates the true posterior, that is:
\begin{align}
    q_{F(\bx, \boldsymbol{\phi})}(\bt) \approx \pr (\bt | \bx).
\end{align}
In general, \acs{SBI} relies on two different strategies to explore the parameter space: amortized posterior estimation or sequential methods. In the former, the neural network is trained once on a dataset generated to cover the entire parameter space. This method is termed `amortized' because, after training, computing the posterior distribution for any mock sample, generated by any set of parameters in the parameter space, is feasible within a matter of seconds. However, this approach requires a substantial number of initial training simulations to cover the whole parameter space, making inferences for a large number of parameters unfeasible. \\
In contrast, sequential methods focus on approximating the posterior distribution $\pr (\bt | \bx_0)$ at a particular observation $\bx_0$. In this case, the learning process is guided by the neural network itself, which focuses on the region of the parameter space that most likely matches the observed data. With this efficient approach, the simulation budget can be considerably reduced, allowing more parameters to be inferred. We note, however, that the amortized nature of the approximate posterior is lost in this case, which is nonetheless only important when having more than one observed data point. In this sense, sequential methods are the most suitable algorithm for this work, where we have a single observed neutron star population and a computationally expensive simulator.\\

\subsection{Truncated sequential neural posterior estimator}
\label{subsec:tsnpe}
\begin{figure*}[t]
\centering
\includegraphics[width = 0.6\textwidth]{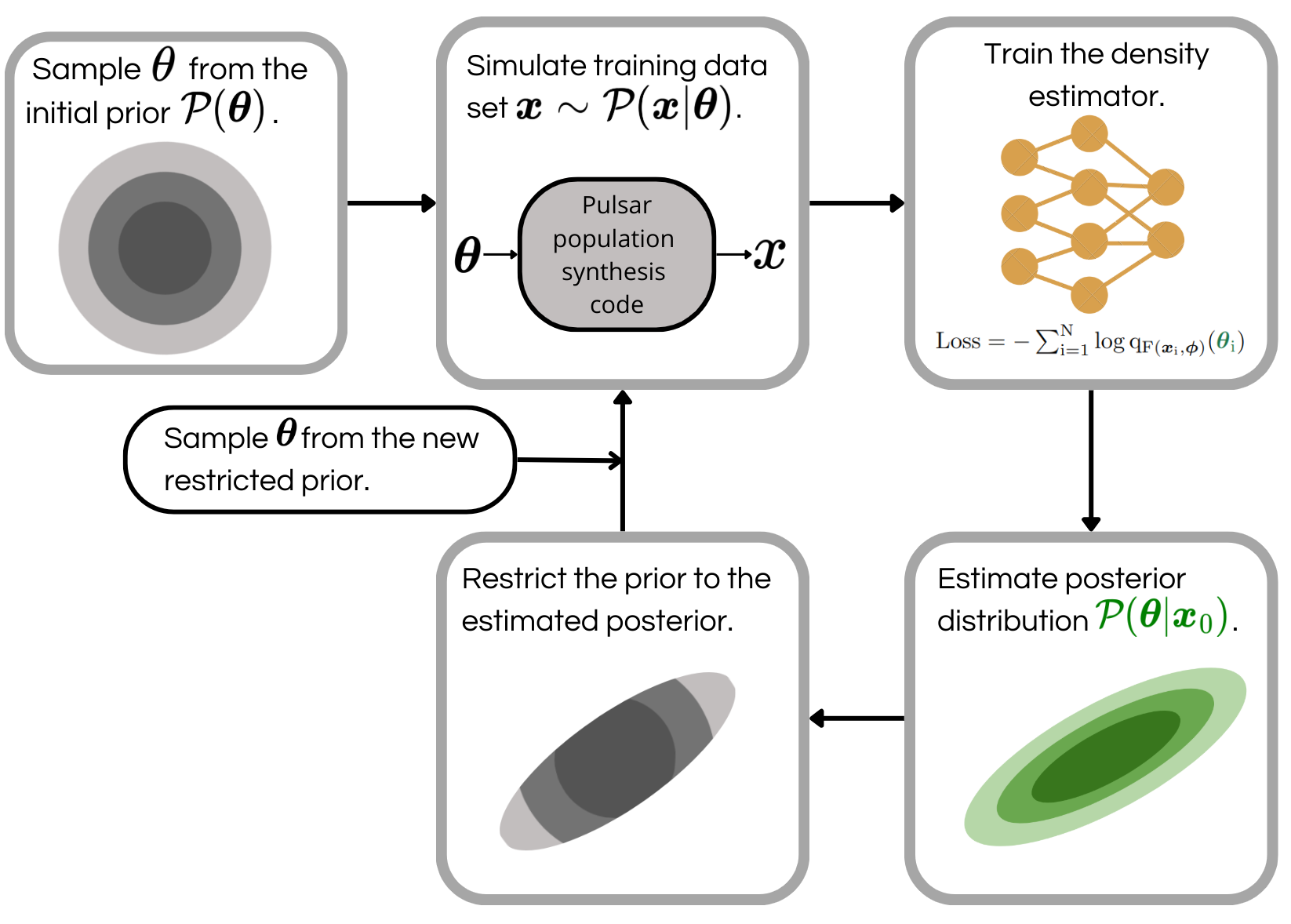}
\caption{Schematic representation of the workflow for the \ac{TSNPE} algorithm applied to pulsar population synthesis.}
\label{fig:workflow_tsnpe}
\end{figure*}
In \citetalias{Graber2024}, we showed how an amortized neural posterior estimator effectively constrains the initial period and magnetic field at birth for the isolated Galactic neutron stars population. However, the number of parameters that we could infer (the five magneto-rotational parameters) was limited due to the high computational cost of producing simulations across the entire parameter space. In this work, we increase the number of inferred parameters to seven by adapting the luminosity law of the Galactic pulsar population (see Equation \ref{eqn:luminosity}). To achieve this, we employ \ac{TSNPE} \citep[][]{Deistler2022}. This approach is significantly more simulation-efficient, as it directs the simulation effort toward the regions of the parameter space that align with the observations. This method is very promising for complex astrophysical simulators and has, for example, been successfully used to perform parameter inference for ringdown gravitational waves \citep{Pacilio2024} and estimate distances of dwarf galaxies \citep{Miller2021}. We summarise the workflow of this algorithm in Figure \ref{fig:workflow_tsnpe}. The \ac{TSNPE} technique uses the following steps:
\begin{enumerate}
    \item Sample the proposal prior distribution to obtain $\bt_i\sim\pr(\bt)$.
    \item Using the simulator, generate synthetic data $\bx_i \sim \pr(\bx|\bt_i)$ based on $\bt_i$ from step 1. 
    \item Train the neural density estimator on the dataset composed of pairs $(\bt_i, \bx_i)$ obtained in the previous steps.
    \item Use the trained neural density estimator to approximate the posterior distribution $\pr (\bt | \bx_0)$ at the observed data, $\bx_0$.
    \item Restrict the prior distribution to the approximated posterior distribution computed in step 4 (see Section \ref{subsec:restrited} below).
    \item Update the proposal prior distribution with the new restricted prior and return to step 1.
\end{enumerate}

We refer to each iteration of this process as `round' and the process continues until the posterior distribution converges. Initially, the proposal prior covers the entire parameter space. However, as the first-round posterior is used to refine the prior in the subsequent round, we do not need to sample as broadly as in \acs{NPE}. The first round provides a rough approximation of the posterior, which removes regions of the parameter space that do not align with the observed data, allowing the subsequent rounds to focus on the more relevant areas. This efficient exploration reduces the total training dataset size significantly. Furthermore, in each round, we train the neural network using the simulations generated from previous rounds combined with those generated in the current round.

\subsection{Restricted prior distribution}
\label{subsec:restrited}
 We denote the restricted prior as $\pr(\bt) \mathbb{1}_{\bt \in \mathcal{M}}$, where $\mathcal{M}$ is the support of the approximated posterior at the observed data, i.e. $q_{F(\bx_0, \boldsymbol{\phi})}(\bt)$, and $\mathbb{1}$ is the indicator function. This means that for all $\bt \in \mathcal{M}$ the restricted prior is equal to the prior while it is 0 otherwise (See Figure \ref{fig:workflow_tsnpe}). We note that the support, $ \mathcal{M}$, refers to the region of the parameter space where the posterior distribution has non-negligible probability mass, essentially indicating the plausible values for the parameters given the observed data. In the following, we restrict the prior to be proportional to the highest-density region of the approximated posterior distribution, which defines the smallest region of the distribution that includes $1-\epsilon$ of the total probability mass. The value of $\epsilon$ is a model hyperparameter, which we fix to $\epsilon = 10^{-4}$ for simplicity, meaning that only $0.01\%$ of the posterior distribution’s support is excluded. To sample the restricted prior distribution and generate parameter samples for the next round, two methods can be used: the rejection method and the Sampling importance resampling (SIR) algorithm \citep{Rubin1988}.\\
In the rejection method, samples are drawn from the prior and accepted if and only if their probability under the approximate posterior is above a certain threshold. However, this approach can lead to a high rejection rate, resulting in unfeasible computation times. In contrast, the SIR algorithm avoids this issue by taking advantage of the relationship
\begin{align}
    \pr(\bt) \mathbb{1}_{\bt \in \mathcal{M}} = \frac{\pr(\bt) \mathbb{1}_{\bt \in \mathcal{M}}}{q_{F(\bx_0, \boldsymbol{\phi})}(\bt)}q_{F(\bx_0, \boldsymbol{\phi})}(\bt)
\end{align} 
and the fact that the approximated posterior distribution $q_{F(\bx_0, \boldsymbol{\phi})}(\bt)$ is easy to sample. The SIR algorithm draws $\mathcal{K}$ samples $\{\bt_1,...,\bt_k\}$ from the approximated posterior distribution and computes their respective weights $ w_k = \frac{\pr(\bt_k) \mathbb{1}_{\bt_k \in \mathcal{M}}}{q_{F(\bx_0, \boldsymbol{\phi})}(\bt_k)}$. These $\mathcal{K}$ samples are then resampled according to the distribution of the normalised weights, $\Tilde{w_k} = \frac{w_k}{\sum_k w_k}$, to obtain $\overline{\bt_m} $ with $ m \in \{1, \dots, M\}$. It can be shown that the cumulative distribution of the normalised weights is equivalent to the cumulative distribution of the restricted prior \citep[See Section 3.2 of][]{Smith92}. Therefore, these new samples $\{\overline{\bt_1},...,\overline{\bt_m}\}$ follow the restricted prior distribution $\pr(\bt) \mathbb{1}_{\bt \in \mathcal{M}}$ when $\mathcal{K} \to \infty$. $\mathcal{K}$ is a hyperparameter of the algorithm. In this work, we follow the suggestion by \cite{Deistler2022} and set $\mathcal{K}=1024$.

\subsection{Approximated posterior validation}
\label{subsec:ppc}
Before evaluating the quality of our posterior approximation at each round, we first need to introduce the concept of coverage probability \citep{Cook2006}.  Coverage probability measures how often the ground truth falls within the estimated \ac{CI} of the posterior distribution for a set of test samples. For example, a $95\%$ \ac{CI} with a $99\%$ coverage probability means that, for $99\%$ of the test samples, the ground truth lies within the $95\%$ \ac{CI} of the approximated posterior distribution. Therefore, the coverage can be used as a metric to assess the quality of the approximated posterior distribution in our test dataset.\\
For a well-calibrated approximated posterior distribution, the coverage probability will match the intended credibility level, resulting in a diagonal line when plotted. A conservative model, however, will have a higher coverage probability and produce wider approximated posterior distributions that contain the true parameter more frequently, thus lying above the diagonal. A conservative estimate, hence, reduces the risk of excluding the ground truth values. In contrast, an overconfident model has a lower coverage probability, resulting in approximated posterior distributions that are too narrow and often miss the ground truth value, placing its coverage probability below the diagonal.\\
To assess our predictions after training, we generate a test dataset using the proposal prior from the current round, combined with the test datasets from all previous rounds. Then, we compute the coverage probability on this combined test dataset to assess the quality of the estimation. Our goal is to achieve a conservative coverage probability, ensuring a broader posterior distribution \citep{Hermans2021}. This strategy helps prevent the exclusion of regions in the parameter space that might match the observed data when restricting the prior using the approximated posterior. 

\subsection{Deep learning setup}
In this work, our neural density estimator consists of a \ac{CNN} combined with a \ac{MDN}, the same network structure as used in \citetalias{Graber2024}.
The CNN takes the density maps described in Section \ref{sec:popsyn} as input and extracts key features into a latent vector of size 32, which is then fed into the \ac{MDN}. The \ac{CNN} architecture includes two 2D-convolutional layers, each followed by a 2D max pooling layer. The \ac{MDN}, a fully connected neural network, outputs the parameters of a Gaussian mixture that approximates the posterior distribution. It contains three fully connected layers with 32 neurons each, followed by an output layer comprising four fully connected sub-layers, corresponding to the mean, weight, diagonal, and upper triangular components of the covariance matrices for the Gaussian mixture. We set the number of components in the mixture to 10, ensuring sufficient flexibility if the posterior is far from a Gaussian distribution.\\

\begin{table*}
\centering
\caption{Summary of the experiments performed in this study. }
\label{tab:experiment_summary}
\begin{tabular}{c c c c c}  
\hline\hline       
\# Exp & Parameters inferred & 1st round sim & Input maps & Inference target \\ 
\hline                    
 1 & 5 mag-rot & 1,000 & 3 $P$-$\dot{P}$ density & Observation\\ 
 2 & 5 mag-rot & 10,000 & 3 $P$-$\dot{P}$ density & Observation \\ 
 3 & 5 mag-rot + 2 lum & 1,000 & 3 $P$-$\dot{P}$ density + 3 $P$-$\dot{P}$ avg flux & Simulation \\ 
 4 & 5 mag-rot + 2 lum & 1,000 & 3 $P$-$\dot{P}$ density + 3 $P$-$\dot{P}$ avg flux & Observation \\ 
 5 & 5 mag-rot  + 2 lum & 10,000 & 3 $P$-$\dot{P}$ density + 3 $P$-$\dot{P}$ avg flux & Observation \\ 
\hline                  
\end{tabular}
\tablefoot{The columns summarise the inferred parameters, the number of training simulations used in the first round, the input maps representing each simulated neutron star population, and the inference targets for each experiment, respectively. For the inferred parameters, we denote those related to the magneto-rotational evolution (i.e., initial period, magnetic field distribution, and late-time magnetic field decay) as `mag-rot' and those related to the luminosity as `lum'. Experiments 1 to 3 are test experiments used to assess the quality of the TSNPE methodology for our pulsar population synthesis.  In bold, we highlight Experiment 4, which provides the main results of this work.}
\end{table*}
Moreover, we fix the batch size to 8 and the learning rate to $5 \times 10^{-4}$. The neural network is trained using the Adam optimiser \citep{Kingma2014}, and we apply an early stopping criterion of 20 epochs to prevent overfitting. This implies that training is stopped if the validation metric does not improve for 20 consecutive epochs, with the best validation weights saved. The weights for the \ac{CNN} are initialized using the Kaiming prescription \citep{Kaiming2015}, while for the \ac{MDN} the weights are initialised with PyTorch's default initialisation. The neural density estimator is implemented using the open-source Python package {\tt sbi} \citep{Tejero-Cantero2020}. The training process is executed on a Tesla V100 SXM2 GPU with $\unit[32]{GB}$ of memory. The generation of simulations in each \ac{TSNPE} round for both the training and test datasets are parallelised to speed up the algorithm. For this, we use the Python package Dask \citep{dask}, a library for dynamic task scheduling. In total, $600$ CPU workers are employed to handle the parallelised simulations. We have tested the robustness of the results presented in Section \ref{sec:results} to variations of these hyper-parameters with additional experiments.

\section{Experiments}
\label{subsec:exps}
We perform two sets of experiments: The first set focuses on testing the \ac{TSNPE} methodology with our pulsar population synthesis, while the second set applies this tested methodology to infer seven parameters (five magneto-rotational parameters and two related to the radio luminosity) for the observed neutron star population. For the testing experiments, we follow two strategies: Experiments 1 and 2 involve inferring the same five magneto-rotational parameters on the observed pulsar population as in \citetalias{Graber2024}: $\mu_{\log B}$, $\sigma_{\log B}$, $\mu_{\log P}$, $\sigma_{\log P}$, and $a_{\rm late}$. We use the results from \citetalias{Graber2024} as a reference for comparison. In test Experiment 3, we expand the parameter space by including two additional luminosity-related parameters, $\mu_{\log L_0}$ and $\alpha$, resulting in a total of seven parameters. In this case, we also add the three $P$-$\dot{P}$ averaged flux maps as input to our neural network. This final test involves inferring these seven parameters for a simulated population with known ground truths, $\bt$. Once we have assessed the quality of our inference technique, we infer the parameters related to the magneto-rotational evolution and the luminosity for the observational data using the fluxes from the TPA program recorded by the MeerKAT telescope (Experiments 4 and 5). All the experiments are summarised in Table \ref{tab:experiment_summary}.

For all experiments, we first generate the training and testing datasets for the first round. Since the initial prior distribution is the same across experiments with the same number of parameters, the training and testing datasets of the first round are shared between the corresponding experiments. For Experiments 1 and 2, we reuse the simulated dataset from \citetalias{Graber2024} and extract a random subsample to match the various sizes of our training and testing datasets for the first round as outlined below. On the other hand, for experiments inferring seven parameters, we generate training and testing datasets of 10,000 and 300 simulations, respectively, using the following uniform priors:
\begin{align}
\label{eqn:priors_7param}
    \mu_{\log B} &\in \mathcal{U}(12, 14), \nonumber \\
    \sigma_{\log B} &\in \mathcal{U}(0.1, 1), \nonumber \\
        \mu_{\log P} & \in \mathcal{U}(-1.5, -0.3), \nonumber \\
    \sigma_{\log P} &\in \mathcal{U}(0.1, 1),   \\
    a_{\rm late} &\in \mathcal{U}(-3, -0.5), \nonumber \\
    \mu_{\log L_0} &\in \mathcal{U}(24.6 , 28.6), \nonumber \\
    \alpha &\in \mathcal{U}(0.1, 1). \nonumber 
\end{align}

The prior ranges for the first five parameters match those used in \citetalias{Graber2024}, while the luminosity parameters are based on the ranges from \citet{Cieslar2020}, although rescaled to match our specific luminosity law in Equation \eqref{eqn:luminosity}. For the experiments constraining five magneto-rotational parameters, we only use the three $P$-$\dot{P}$ maps as in \citetalias{Graber2024} as an input (shown in the top row of Figure \ref{fig:density_maps_test}). On the other hand, when inferring the parameters related to the luminosity together with those related to the magneto-rotational evolution in Experiments 3 to 5, we add the three $P$-$\dot{P}$ averaged flux maps (see bottom row of Figure \ref{fig:density_maps_test}). As a result, in this case the input for the neural network consists of six maps.

To manage the computational expenses of generating new simulations from the restricted priors in subsequent rounds, we fix the number of simulations generated in each round for the training dataset. From rounds 2 through 10, the newly generated simulations for the training dataset are fixed at 1,000 simulations, and for the test dataset at 300 simulations. This setup provides sufficient data to train the model and evaluate the coverage probability in each round while maintaining manageable computational costs. In each round, $10\%$ of the training dataset is used for validation. 

All experiments are run for 10 rounds allowing us to track how the posterior distributions evolve across the rounds. Additionally, we train the neural networks from scratch in each round, i.e., the weights are randomly initialized at the start of every round. We opt for this approach because we observe that using pre-trained weights from previous rounds tends to result in overconfident posterior approximations. For each round, we train five independent neural networks with identical hyperparameters and architecture on the same training dataset. The only difference between them is the randomly initialized weights. The predictions from these five networks are then combined to form an ensemble posterior distribution \citepalias[see Section 4.2 of][for details on the ensemble]{Graber2024}. We adopt this approach because, when using a single neural network, the coverage probability suggests the posterior distribution to be overconfident, i.e., the approximated posterior is narrower than the true distribution (see Section \ref{subsec:ppc} for details on coverage probability). By using an ensemble, we avoid restricting the prior distribution too much, thus preventing the exclusion of relevant regions in the parameter space \citep{Hermans2019}.

\begin{figure*}[t!]
\centering
\includegraphics[width = 1\textwidth]{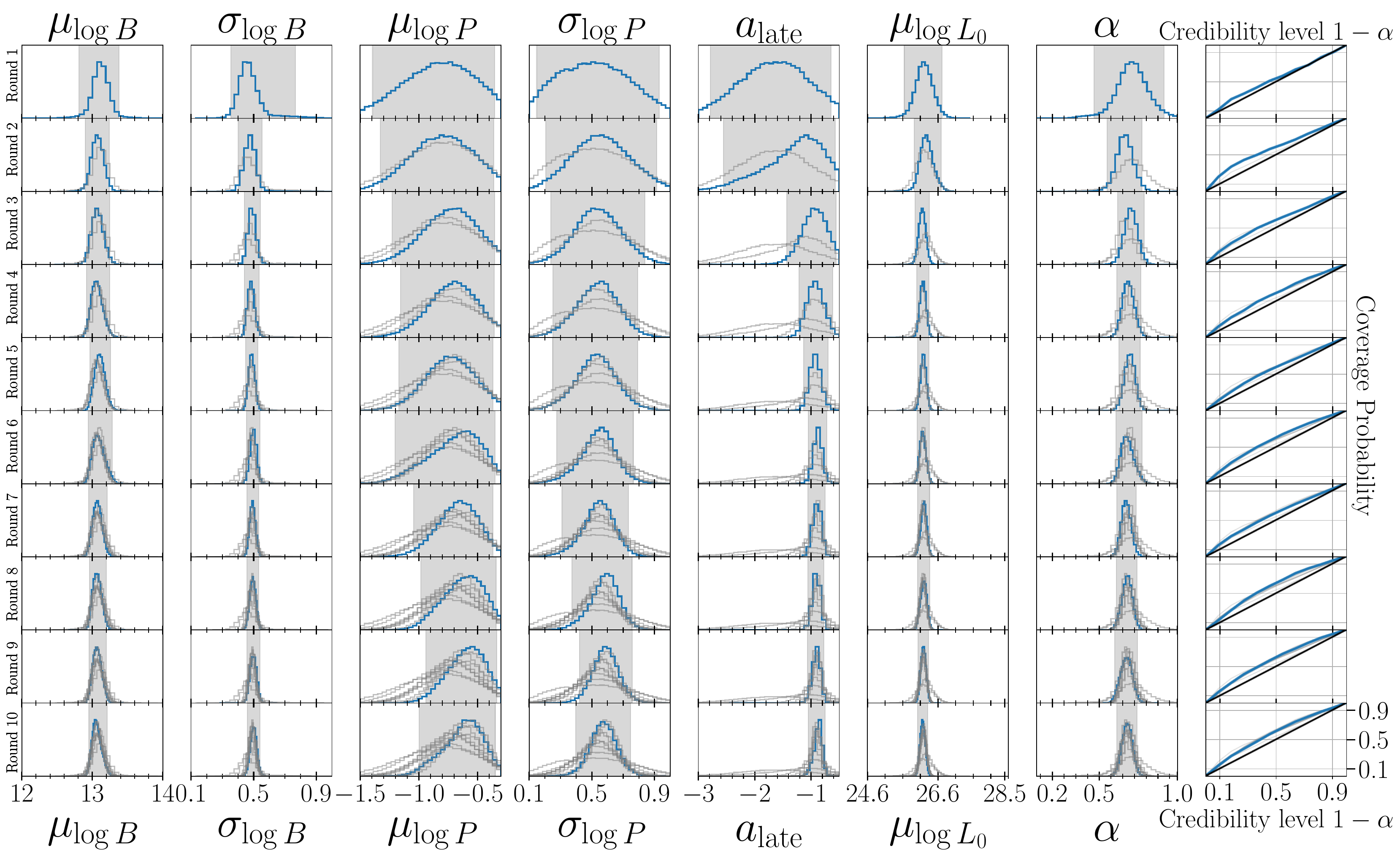}
\caption{Illustration of the \ac{TSNPE} algorithm applied to our pulsar population synthesis. Here, we show the results for inferring the seven free parameters related to the magneto-rotational evolution and the luminosity for the observed neutron star population, using 1,000 simulations in the first round in Experiment 4. Each row corresponds to one round of inference. The last column shows the coverage probability computed on the test dataset. In each panel, the current round's computed values are shown in blue, while values from previous rounds are shown in light grey. The grey shaded area in the one-dimensional marginal posterior represents the $95\%$ credibility interval of the approximated posterior for that round. In each of these 1D marginal posterior distributions, the horizontal axes represent the parameters’ prior ranges. The results from Experiment 5 are qualitatively similar to those presented here.}
\label{fig:exp_11}
\end{figure*}

\section{Results}
\label{sec:results}

\subsection{Five-parameter test: Magneto-rotational}
\label{subsec:test-5}

We first assess the performance of the TSNPE algorithm (explained in Section \ref{subsec:tsnpe}) when inferring the same number of parameters as in \citetalias{Graber2024} to test its performance. We use the results of \citetalias{Graber2024} as a benchmark for this five-parameter experiments. Figures \ref{fig:exp_1}
and \ref{fig:exp_2} in Appendix \ref{sec:appendix} show the results of the Experiments 1 and 2, respectively. 
Each row displays the 1D marginal posterior distribution for each parameter for a given round, with the coverage probability shown in the last column. We note that the $x$-axis range in each of the 1D marginal posterior distributions is the initial prior range. \\

In Experiment 1, where we use 1,000 simulations for the first round, the approximated posterior distributions (shown in blue in Figure \ref{fig:exp_1}) for the period parameters ($\mu_{\log P}, \sigma_{\log P}$) are broad. The posterior support spans a range that is similar to the prior distribution. In contrast, for the magnetic field parameters ($\mu_{\log B}, \sigma_{\log B}, a_{\rm late}$), the first round already narrows down a significant portion of the parameter space allowing subsequent rounds to achieve better approximations and tighter posterior distributions. While the approximated posterior distribution for the initial magnetic field quickly converges after round 2, the inferred posterior for the initial period and magnetic field decay at late times does not appear to converge to a fixed distribution. Instead, the distributions for these parameters shift slightly from one round to the next and no longer seem to narrow further after round 8. Additionally, examining the coverage probability, we see that it closely follows slightly above the diagonal, indicating a conservative estimate of the posterior distribution for all rounds that does not improve significantly as the \ac{TSNPE} progresses. We associate this with the fact that the test dataset in each round includes simulations generated in that round, along with those from previous rounds. The inferred posterior distributions for the magnetic field at birth agree with those approximated in \citetalias{Graber2024} (shown in black). This is not the case for the approximated posteriors for the initial period and magnetic field decay at late times as they are shifted compared to the estimates of \citetalias{Graber2024}. We however note that these posteriors are also approximations obtained using \ac{NPE} and should not be taken as the ground truth. \\

Given these observed shifts in some of the parameters, we also test the effect of increasing the training dataset size in the first round to 10,000 simulations in Experiment 2. The corresponding results are shown in Figure \ref{fig:exp_2}. In the first round, the posterior distributions for all parameters are already well constrained. While in this initial round the approximated posteriors for the initial magnetic field align with those of \citetalias{Graber2024}, the initial period approximated posteriors still show slight shifts compared to our earlier results. However, as the rounds progress, the inferred posterior gradually aligns more closely with the results from \citetalias{Graber2024}. Additionally, by round 7, the parameter $a_{\rm late}$ begins to exhibit bimodality, similar to what was observed in our previous study. This bimodality arises because the individual neural networks within the ensemble produce different estimates for this parameter, as already pointed out in Section 5.5 of \citetalias{Graber2024}. We also observe some oscillations in the $\sigma_{\log P}$ and $a_{\rm late}$ approximated posteriors between a narrower distribution and one that aligns more closely with that of \citetalias{Graber2024}. Despite these variations, after five rounds, the posterior distributions largely agree with the estimates from \citetalias{Graber2024}. Moreover, the coverage probability moves closer to the diagonal as the rounds progress, indicating that the posterior is becoming more accurate. This demonstrates that \ac{TSNPE} requires only about 19,000 training simulations to obtain well constrained posterior distributions, compared to the 360,000 simulations used in \citetalias{Graber2024}, highlighting the efficiency of \ac{TSNPE} over \ac{NPE} for this problem.

\subsection{Seven-parameter test: Magneto-rotational and luminosity}
\label{subsec:test-7}
We next apply the \ac{TSNPE} method to estimate a set of seven parameters: the same five magneto-rotational parameters used in \citetalias{Graber2024} along with two additional parameters related to the bolometric intrinsic radio luminosity law of Equation \eqref{eqn:luminosity}. As a further test of our method, we first performed Experiment 3 to evaluate the quality of the approximated posterior distribution. In this experiment, the \ac{TSNPE} algorithm is focused on inferring the posterior distribution of a simulated population of neutron stars with known ground truth parameters $\bt$. \\
The results of Experiment 3 are presented in Figure \ref{fig:exp_3} in Appendix \ref{sec:appendix}. Similar to previous experiments, each row depicts the 1D marginal posterior distribution for each parameter, with the last column showing the coverage probability for the test dataset. In Figure \ref{fig:exp_3}, we mark the ground truth parameters, $\bt$, by orange dashed lines.\\

During the first round the posterior approximations for the initial period parameters ($\mu_{\log P}, \sigma_{\log P} $) and the magnetic field decay at late times parameter ($a_{\rm late}$) remain as broad as the prior ranges. For the parameters related to the initial magnetic field distribution ($\mu_{\log B}, \sigma_{\log B}$) and the luminosity law ($\mu_{\log L_0}, \alpha$), the algorithm already discards significant portions of the parameter space in this first round, enabling more refined approximations in subsequent rounds. We highlight that the true value is not necessarily expected to coincide with the peak of the posterior distribution. By definition, a well-calibrated posterior distribution should have coverage that aligns with the diagonal, meaning that, for instance, $99\%$ of the ground truth values will fall within the $99\%$ \ac{CI} of the respective approximated posteriors. Therefore, while the ground truth may not match the peak of the posterior, it should lie within the support of the approximated posterior. In Figure \ref{fig:exp_3}, the ground truth falls within the $95\%$ \ac{CI} (shaded in gray), demonstrating that \ac{TSNPE} successfully recovers the posterior distribution for the simulated neutron star population when using a total of 1,000 simulations. \\

We further observe that as the rounds progress, the coverage stays above the diagonal, suggesting that the posterior estimates are slightly conservative, without significant changes as we iterate through the \ac{TSNPE} algorithm. For the $\sigma_{\log B}$ parameter, an interesting phenomenon occurs in the early rounds. While in the first round, the approximated posterior distribution is shifted relative to the ground truth, in the second round, a secondary peak forms close to this value. By the third round, the neural network appears to favor this secondary peak, which aligns with the ground truth. In subsequent rounds, the approximated posterior converges to the posterior from this round, highlighting the efficiency of the algorithm.  \\

\subsection{Magneto-rotational and luminosity estimation inferred from the observed population}
\label{subsec:main_results}
As we have demonstrated the effectiveness of \ac{TSNPE} when applied to our pulsar population synthesis, we now turn our attention to the main results of this work where we infer the parameters related to the luminosity and the magneto-rotational evolution for the observed population in Experiments 4 and 5. The difference between these experiments is the number of simulations in the first round (see Table \ref{tab:experiment_summary}). As we observe no significant differences in the inference results between them, we only present the results of Experiment 4 in Figure \ref{fig:exp_11}. The consistency between both experiments increases our confidence in the robustness of the inference results.

In the first round of Experiment 4, as in Experiment 3, the 1D marginals of the approximated posterior distribution for the parameters $\mu_{\log P}, \sigma_{\log P}$ and $a_{\rm late}$ remain as broad as the prior ranges. However, for $\mu_{\log B}$, $\sigma_{\log B}$, $\mu_{\log L_0}$ and $\alpha$, a wide range of the parameter space is removed, which helps better constrain the initial period distribution and the late-time magnetic field decay in subsequent rounds. 
\begin{figure*}
\centering
\includegraphics[width = 0.8\textwidth]{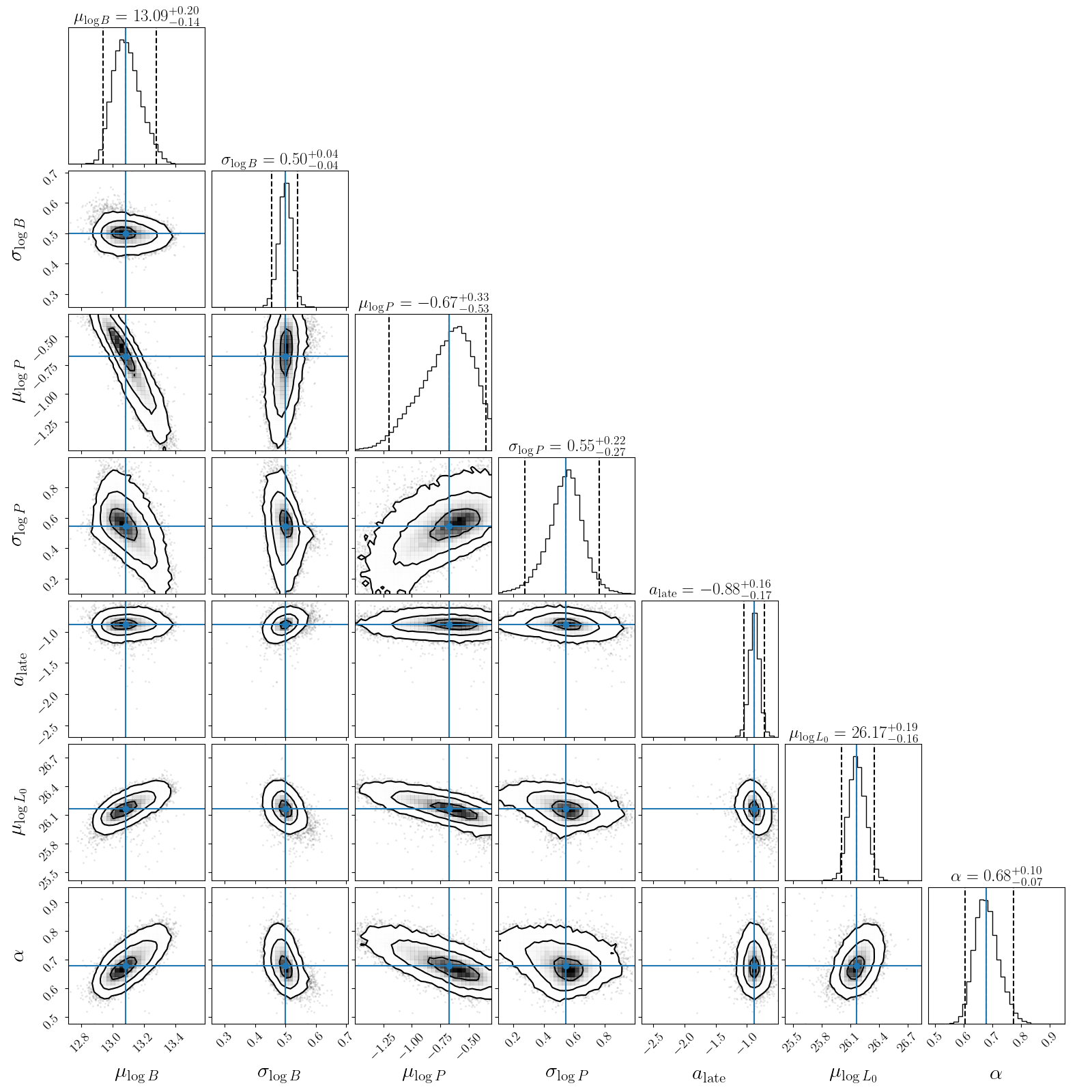}
\caption{Inference results for the observed pulsar population using round 6 of Experiment 4. The corner plot shows 1D and 2D marginal posterior distributions for the five magneto-rotational parameters and the two parameters related to the intrinsic bolometric luminosity. We highlight the medians in light blue. Corresponding values and $95\%$ \acp{CI} are summarized above the panels and in Equation \eqref{eqn:CI_best}.}
\label{fig:exp_11_corner_plot}
\end{figure*}
\begin{table*}
\caption{Comparison between best parameters for the log-normal initial magnetic-field, initial period distributions, and the magnetic field decay at late times in the literature.}    
\label{tab:infer_comparison}      
\centering  
\renewcommand{\arraystretch}{1.5} 
\begin{tabular}{c@{\hskip 5pt}c@{\hskip 5pt}c@{\hskip 5pt}c@{\hskip 5pt}c@{\hskip 5pt}c@{\hskip 5pt}c@{\hskip 5pt}c}  
\hline\hline       
References & $\mu_{\log B}$ & $\sigma_{\log B}$ & $\mu_{\log P}$ & $\sigma_{\log P}$ & $a_{\text{late}}$ & $\mu_{\log L_0}$ & $\alpha$\\ 
\hline                    
\citet{Faucher2006} & $12.65$ & $0.55$ & ... & ... & ... & ... & ... \\
\citet{Gullon2015}  & $12.99$ & $0.56$ & ... & ... & ... & ... & ...\\
\citet{Cieslar2020} & $12.67^{+0.01}_{-0.02}$ & $0.34^{+0.02}_{-0.01}$ & ... & ... & ... \\
\citet{Igoshev2022} & $12.44$ & $0.44$ & $-1.04^{+0.15}_{-0.20}$ & $0.53^{+0.12}_{-0.08}$ & ... & ... & ... \\
\citet{Sautron} & $12.44$ & $0.5$ & $-0.88$ & $0.45$ & ... & ... & ... \\
\citet{Zhihong2024} & $12.32^{+0.05}_{-0.07}$ & $0.35^{+0.03}_{-0.03}$ & $-1.40^{+0.48}_{-1.30}$ & $0.35^{+0.84}_{-0.35}$ & ... & ... & ... \\
\citetalias{Graber2024} & $13.10^{+0.04}_{-0.05}$ & $0.45^{+0.03}_{-0.02}$ & $-1.00^{+0.11}_{-0.10}$ & $0.38^{+0.16}_{-0.10}$ & $-1.80^{+0.65}_{-0.61}$ & ... & ... \\
This work & $13.09^{+0.10}_{-0.08}$ &  $0.50^{+0.02}_{-0.02}$ & $-0.67^{+0.20}_{-0.27}$ & $0.55^{+0.10}_{-0.12}$ & $-0.88^{+0.08}_{-0.08}$ & $26.17^{+0.09}_{-0.08}$ & $0.68^{+0.05}_{-0.04}$ \\
\hline\hline
\end{tabular}
\tablefoot{We provide references and the seven relevant parameters. We note that the first three studies use a different prescription for the initial period, which prevents a direct comparison with our study. Additionally, the $a_{\rm late}$ parameter was first introduced in \citetalias{Graber2024}, so the comparison can only be made with that work. Regarding the parameters related to the luminosity, the prescription used in this work differs from that in previous studies, making direct comparisons not possible. For \citet{Gullon2015}, \citet{Cieslar2020} and \citet{Zhihong2024}, we compare with "model D" for the radio-pulsar population, the "rotational model" and the exponential decay model for the magnetic field, respectively. Where available, we quote \acp{CI} at the $68\%$ level (including for this work), but we note that these are difficult to compare due to the difference in inference methods and underlying models and data.}
\end{table*}

For all the parameters with the exception of $\mu_{\log P}$ and $\sigma_{\log P}$, the approximated marginal posterior distributions converge after round 5. On the other hand, for the initial period parameters, we observe a small shift in the left tail of the marginal posteriors from one round to the next, beginning in round 7. Although the shift is small, we conservatively use the results from round 6 as the best estimate. At this point, the marginal posterior distribution is sufficiently broad to cover the range of values observed in rounds 7 to 10.  A corner plot for the 1D and 2D  posteriors of round 6 is illustrated
in Figure \ref{fig:exp_11_corner_plot}. Corresponding medians are shown in light blue. In the 2D marginal posterior distribution, we observe correlations between several parameters. For this reason, we compute the Pearson and Spearman correlation coefficients and consider parameters to be correlated if and only if their absolute value is greater than $0.5$ and the corresponding p-value $<0.05$. As a result, we observed strong correlations between the parameters related to the luminosity and the initial period as well as the magnetic field distributions. We discuss the physical interpretations of these correlations in detail below.

In contrast to our results in \citetalias{Graber2024} and Experiments 1 and 2, the $a_{\rm late}$ parameter appears to be well constrained and does not exhibit any bimodality or shifts between rounds. We will further discuss our interpretation of this improvement in the next section. Finally, in the rightmost panels of Figure \ref{fig:exp_11}, we observe the coverage probability to remain above the diagonal for all rounds without significant changes between them, indicating a conservative estimate of the approximated posterior distribution.

\section{Discussion}
\label{sec:conclusions}
\subsection{Inference technique: TSNPE}
We begin our discussion by comparing the results of this work based on \ac{TSNPE} with those obtained in \citetalias{Graber2024} using \ac{NPE}. While for the initial magnetic field the estimated distributions match the estimates of \citetalias{Graber2024}, the new inferred distributions for the initial period and magnetic field decay at late times differ slightly from these previous estimates. These differences are more pronounced when we use $1,000$ simulations in the first round (see Figure \ref{fig:exp_1}). However, when the first round training dataset size is increased to $10,000$ simulations, the approximated posteriors move closer to the previous estimates, and by round 10, they closely align with those in \citetalias{Graber2024} (See Figure \ref{fig:exp_2}). This behaviour is expected as the network is trained with more simulations and thus gains better insight into the parameter landscape, allowing the network to more accurately approximate the posterior distribution. However, when inferring seven parameters for the observed population, the results obtained by round 10 using $10,000$ simulations for the training in the first round are comparable to those obtained with only $1,000$ initial training simulations. This suggests that the $1,000$ training simulations used in the first round are already sufficient for the neural network to estimate the seven parameters, and adding more does not provide additional information for this task. As discussed below in detail, we associate this improvement with the addition of flux measurements as inputs to the neural density estimator.

The broader marginal posterior distributions for the initial period, compared to the narrower constraints on the other parameters, suggest a degeneracy in the initial period. For example, in Figure \ref{fig:exp_11} where we infer the seven parameters for the observed data, the distribution for the initial magnetic field in the first round is already very narrow. In contrast, the distribution for the initial period remains relatively broad even in the final round, showing a slight shift across rounds. This suggests that estimating $\mu_{\log P}$ and $\sigma_{\log P}$ is a challenge, as already noted in earlier works using different inference methods \citep{Gullon2014,Graber2024}. As previously discussed in Section 5.3 of \citetalias{Graber2024}, the initial period information is gradually lost during the evolution of neutron stars in the $P$-$\dot{P}$ diagram. This effect is easily visualized when the magnetic field is constant, although similar reasoning applies when decay is introduced: Two neutron stars with different initial periods but the same initial magnetic field at birth lie on the same constant magnetic field line in the $P$-$\dot{P}$ diagram. Although both sources move along this constant field line throughout their evolution, the one with a shorter period evolves faster than the one with a longer period (see Equation \ref{eqn:P_ode}). Therefore, by the end of the simulation, both sources end up in similar locations in the $P$-$\dot{P}$ plane, making it difficult to distinguish between them. This degeneracy effectively leads to broad posterior distributions for the initial period as observed.
\subsection{Magnetic field decay at late times}

Now we turn our attention to the magnetic field decay at late times. When we infer seven parameters and add the flux measurement as input, this parameter is narrowly constrained, indicating that the density estimator is confident in the inference of this parameter. This is evident in the fifth panel of Figure \ref{fig:exp_11}. This is in contrast to what we observe in \citetalias{Graber2024} and in the testing experiments where we infer five parameters, which estimate a broader and bimodal distribution. Bimodality in the posterior distribution for this parameter arises when the different neural networks in the ensemble disagree, leading to non-overlapping supports.

To investigate the reason for the improved $a_{\rm late}$ constraints, we focus on the two main changes in this work beyond the \ac{TSNPE} implementation: i) modifying our luminosity prescription (see Equation \ref{eqn:luminosity}) and adding two additional free parameters and ii) providing additional $P$-$\dot{P}$ averaged flux maps. The former is relevant because the bolometric intrinsic luminosity is proportional to the loss of the rotational energy and the dipolar magnetic field and luminosity are thus related. Therefore, the luminosity of old neutron stars (i.e., $t_{\rm age} > \tau_{\rm late}$) is influenced by the $a_{\rm late}$ value. This effect should primarily affect the synthetic \ac{SMPS} populations which focus on high latitudes and contain a larger fraction of old pulsars with respect to young pulsars compared to \ac{PMPS} and \ac{HTRU}. Specifically, for two simulations with identical parameters but different $a_{\rm late}$ values, the $P$-$\dot{P}$ averaged flux maps for \ac{SMPS} can differ significantly while \ac{PMPS} and \ac{HTRU} are unaffected. We explore this effect on our inference by computing the correlation between $a_{\rm late}$ and those parameters related to the luminosity in the 2D posteriors in Figure \ref{fig:exp_11_corner_plot}. We find that both the Pearson and Spearman coefficients are lower than 0.5, indicating no significant correlation between these two variables. We emphasise, however, that the influence of $a_{\rm late}$ is primarily on the \ac{SMPS}, while the correlation is computed across all three surveys. The coefficients are thus not overly sensitive to this correlation. Additionally, we note that the luminosity prescription used when constraining seven parameters differs slightly from that applied when inferring only five, where the prescription from \citetalias{Graber2024} was applied. Nonetheless, we cannot associate the new luminosity prescription solely with the improved $a_{\rm late}$ constraints presented in this work.
 
 Therefore, we now turn our attention to the effect of adding the $P$-$\dot{P}$ averaged flux maps as input to the neural density estimator. To test this, we perform an experiment where we infer the seven parameters providing only the three $P$-$\dot{P}$ density maps. We observe that the $a_{\rm late}$ posterior becomes broader, and bimodality arises, highlighting the importance of the flux measurements in the inference process. Additionally, this experiment shows that the approximated posterior distributions for parameters related to the initial period and luminosity become broader as well. Providing the density estimator with additional information on the fluxes of the pulsar population when inferring the magnetic field at late times turns out to be crucial and enhances the overall inference accuracy.
 
\subsection{Best estimated parameters with TSNPE }
After assessing the quality of the \ac{TSNPE} algorithm with the testing experiments, we used this technique to infer the magneto-rotational luminosity parameters for the observed population. In particular, we chose as our `best estimates' the values from round 6 of Experiment 4 as outlined above in Section \ref{subsec:main_results}. Using the corresponding density estimator, we find the following best estimates at $95\%$ credible level:
\begin{align}
	\mu_{\log B} &= 13.09^{+0.20}_{-0.14}, \nonumber \\
	\sigma_{\log B} &= 0.50^{+0.04}_{-0.04}, \nonumber \\
	\mu_{\log P} &= -0.67^{+0.33}_{-0.53}, \nonumber \\
	\sigma_{\log P} &= 0.55^{+0.22}_{-0.27}, \label{eqn:CI_best} \\
	a_{\rm late} &= -0.88^{+0.16}_{-0.17}, \nonumber \\
        \mu_{\log L_0} &= 26.17^{+0.19}_{-0.16}, \nonumber \\
        \alpha &= 0.68^{+0.10}_{-0.07}. \nonumber
\end{align}

From the Pearson and Spearman correlation coefficients calculated for the 2D marginals shown in Figure \ref{fig:exp_11_corner_plot}, we find a positive correlation between $\mu_{\log L_0}$ and $\mu_{\log B_0}$, which indicates that higher initial magnetic fields require an increase in $\mu_{\log L_0}$ to fit the observed data. This relationship is expected because a higher $B_0$ shifts the $\dot{P}$ distribution of detected pulsars towards higher values. To match the observations, the model requires an increase in luminosity facilitating the detection of pulsars with lower $\dot{P}$. We also observe a negative correlation between $\mu_{\log L_0}$ and $\mu_{\log P_0}$. Therefore, a longer initial period distribution leads to a lower $\mu_{\log L_0}$. Increasing $\mu_{\log L_0}$ shifts the simulated population to the right in the $P$-$\dot{P}$ diagram since pulsars with lower rotational energy become detectable. This shift requires the population to move toward shorter final periods to align the simulated population with observations. This, in turn, can be compensated by a shift towards shorter periods, resulting in the observed anti-correlation. \\


\begin{figure*}

\centering
\includegraphics[height=0.6\columnwidth]{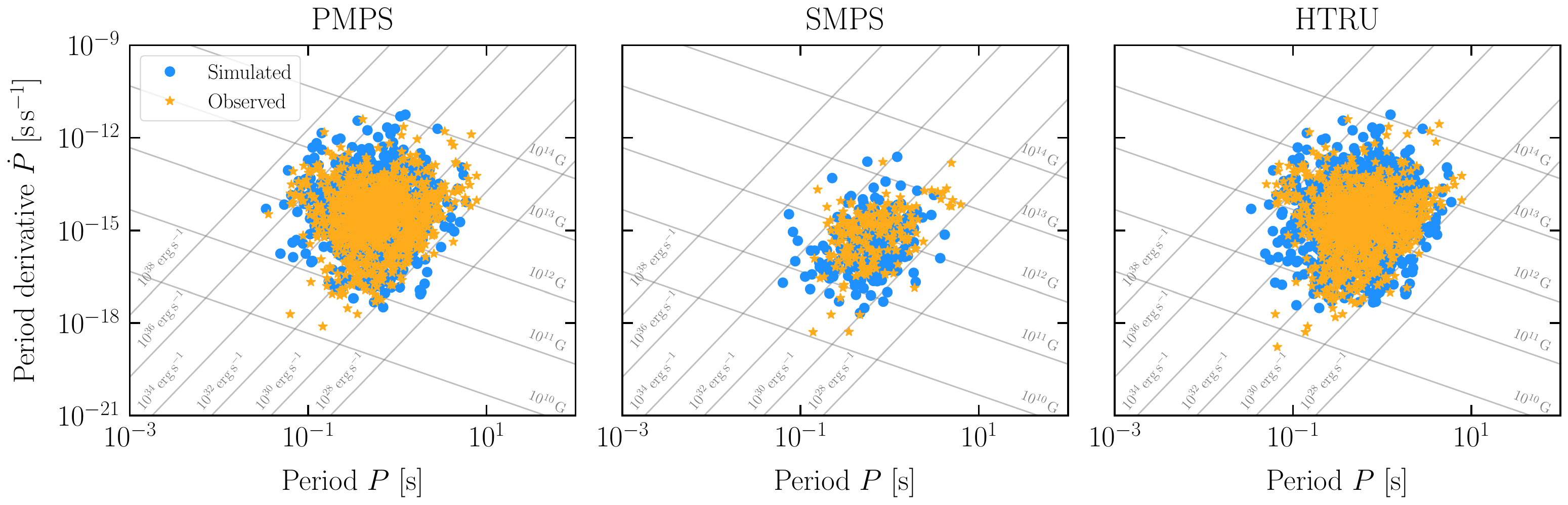}
\caption{Simulated and observed populations of isolated Galactic radio pulsars. Each panel corresponds to a different survey, from left to right: PMPS, SMPS, and the low- and mid-latitude HTRU survey. The yellow stars indicate the observed pulsar population with data taken from the ATNF Pulsar Catalogue \citep[][v2.5.1]{Manchester2005}. The blue dots represent the simulated pulsar population for the parameters inferred via \ac{TSNPE} (see Equation \ref{eqn:CI_best}). Lines of constant spin-down power ($|\dot{E}_{\rm rot}|$) and constant dipolar surface magnetic field are also shown.}
	\label{fig:ppdots_obs_vs_sim}
\end{figure*}

\begin{figure}
	\centering
	\includegraphics[width=0.85\columnwidth]{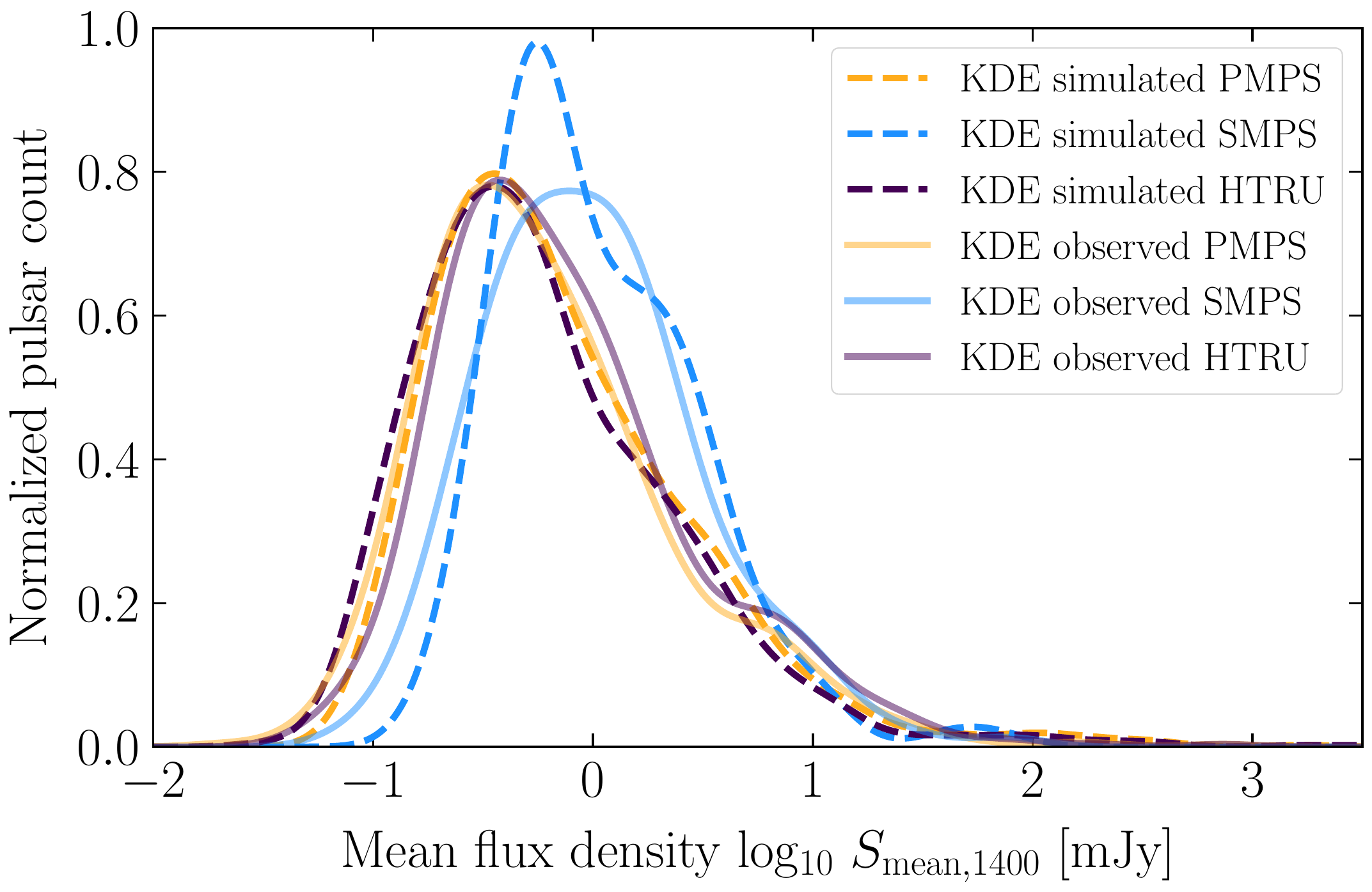}
    \caption{Distributions of mean radio flux densities, $S_{{\rm mean}, 1400}$, at \unit[1400]{MHz} for the populations of isolated Galactic radio pulsars in the \acs{PMPS}, the \acs{SMPS}, and the low- and mid-latitude \acs{HTRU} survey (in yellow, light blue and purple, respectively). We show the individual probability density functions obtained via \acs{KDE} using a Gaussian kernel to facilitate comparison. Estimates for the observed population are shown as solid lines, while our best-parameter simulation is shown with dashed lines. Data taken from the TPA program \citep[][]{Posselt2023}.}
	\label{fig:flux_KDE_comp}
\end{figure}

Finally, we compared the simulated population using the best-estimated parameters shown in Equation \eqref{eqn:CI_best} and the observed population. Although an exhaustive comparison is beyond the scope of this work, we run a simulation with these best estimates and provide Figures \ref{fig:ppdots_obs_vs_sim} and \ref{fig:flux_KDE_comp} to discuss the main aspects. In Figure \ref{fig:ppdots_obs_vs_sim}, we show the $P$-$\dot{P}$ maps for both the observed and the simulated neutron star populations for each of the surveys. The simulated pulsar population closely resembles the observed population. Notably, in \citetalias{Graber2024}, the \ac{SMPS} sample showed a slight shift towards lower $\dot{P}$ values, which we attributed to the weak constraints on $a_{\rm late}$ parameter. We do not observe the same shift in this work as we more tightly constrain this value with the \ac{TSNPE} algorithm. Overall, the distributions of the best-estimated simulated neutron star population closely resemble those of the observed population, giving us confidence in our inference results. Moreover, in Figure \ref{fig:flux_KDE_comp}, we show the kernel density estimation (KDE) fits for the flux density distribution obtained with a Gaussian kernel, to easily compare between the observed and simulated populations. While the distributions for both \ac{PMPS} and \ac{HTRU} closely follow those of the observed population, the simulated fluxes for \ac{SMPS} differ slightly from the observed ones. This discrepancy might hint at missing physics at late times because (as we already mentioned) \ac{SMPS} is more sensitive to old pulsars compared to the other two surveys. 

To compute the inferred birth rate for each survey, we run 10 different simulations using the best-parameter estimates. The resulting means and standard deviation for the birthrates are as follows:

\begin{align}
&\text{\ac{PMPS}: $\sim 2.16 \pm 0.09$ neutron stars per century}, \nonumber \\
&\text{\ac{SMPS}: $\sim 1.9 \pm 0.09$ neutron stars per century}, \label{eqn:BR_estimated} \\
&\text{\ac{HTRU}: $\sim 1.7 \pm 0.07$ neutron stars per century}, \nonumber
\end{align}

\noindent These estimates are comparable to those found in \citetalias{Graber2024} and compatible with the recent core-collapse supernova rate inferred by \citet{Rozwadowska2021}.

Finally, Table \ref{tab:infer_comparison} compares the results of this work with previous studies. The inferred initial magnetic field distribution is consistent with prior estimates that apply a similarly realistic prescription for the magnetic field decay, as in the work of \cite{Gullon2014} and in \citetalias{Graber2024}. However, the initial period distribution tends toward slightly larger values compared to \citetalias{Graber2024} and previous works. In the former, this difference may be explained by the fact that we fix the value of $\mu_{\log L_0}$ to a higher value in \citetalias{Graber2024} than estimated here. The anti-correlation between $\mu_{\log L_0}$ and $\mu_{\log P_0}$, therefore, biases the inference of the initial period distribution toward shorter periods. On the other hand, \cite{Igoshev2022} analysed 56 young neutron stars associated with supernova remnants and studied their magneto-rotational properties only. Their estimates may be biased due to the small number of relatively young sources in their analysis, which could explain the discrepancy between their work and this one. We further note that our results cannot be directly compared with those of \cite{Sautron} as their study does not include parameter estimation and adopts a pseudo-luminosity prescription instead of the intrinsic bolometric luminosity used here. On the other hand, the relationship we found between the luminosity, $P$, and $\dot{P}$ is consistent, within uncertainties, with the results of \cite{Zhihong2024}.
However, their inferred normalisation factor, $L_0$, differs considerably from ours, likely due to a completely different beaming prescription. Additionally, both \cite{Sautron} and \cite{Zhihong2024} employ magnetic field decay prescriptions that differ significantly from the one used in this work, which additionally contributes to the observed discrepancies in the inferred parameters.

\section{Summary}

 In this work, we demonstrate how combining a sequential \acs{SBI} approach with a pulsar population synthesis framework together with consistent flux measurement allows us to successfully infer the parameters related to magneto-rotational evolution. This new approach also sheds light on the distribution of the intrinsic bolometric radio luminosity of the Galactic isolated neutron star population. Our main findings are as follows:\\
\begin{itemize}
    
    \item When inferring the parameters related to the luminosity and the magneto-rotational evolution, we include flux measurements as an additional input. Specifically, for the observed neutron star population, we use data from the TPA program recorded by the MeerKAT telescope, as reported by \cite{Posselt2023}. This dataset is the largest unified neutron star catalogue with consistent flux measurement. Incorporating flux information not only helps us constrain the luminosity but also leads to tighter constraints on the late-time magnetic field decay, compared to our previous work.

    \item We successfully recover narrow posterior distributions for the parameters related to the magnetic field and the luminosity. The inferred period distribution is broader compared to the rest of the parameters. This is expected due to the degeneracies between the period and already noted in previous works using different inference techniques. Overall, the approximated posterior distributions obtained are consistent with previous studies.
    
    \item While our previous study using NPE (see \citetalias{Graber2024}) required 360,000 training simulations to estimate the five magneto-rotational parameters, here we achieve comparable results with only 19,000 simulations due to the efficiency of the sequential technique (see Figure \ref{fig:exp_2}). This efficiency enables us to expand the number of inferrable parameters which was unfeasible with the \ac{NPE} method.
\end{itemize}

As mentioned previously and also noted in \citetalias{Graber2024} and other works, the posterior distribution for the initial period is difficult to constrain, resulting in a broader distribution. This is because the information about the initial period is lost during evolution, making it challenging for any inference approach to estimate these parameters. New pulsar surveys in the radio band, as well as in other wavelengths, will increase the number of observed neutron stars, which may help constrain the neutron star population further. Moreover, it is crucial to include the entire pulsar population, not just radio pulsars. As shown in \cite{Gullon2015}, including magnetars, highly magnetized neutron stars emitting X-rays, help constrain the higher end of the magnetic field distribution at birth. Current population synthesis models are naturally biased toward lower magnetic field neutron stars. \\
Furthermore, as reflected in the discrepancies between the simulated flux distribution for the \ac{SMPS} survey and the observed data, we are still missing some relevant physics in the late-time evolution. Comparing different prescriptions for magnetic field decay at late times, will be crucial in the future. Having demonstrated that \ac{TSNPE} can successfully infer multiple parameters of the isolated pulsar population synthesis with a reduced simulation budget. Model comparison with \acs{SBI} as e.g. introduced in \cite{Mancini2024} is now within reach to help us deepen our understanding of the evolution of old pulsars further.

\begin{acknowledgements}
The authors thank Emilie Parent for useful exchanges on radio pulsar emission and detections, Clara Dehman for providing magnetic field evolution curves, Bettina Posselt and Aris Karastergiou for useful discussion and insights on the TPA program data, and Michael Deistler and Maximilian Dax for valuable discussions on SBI and TSNPE. The data production, processing, and analysis tools for this paper have been implemented and operated at the Port d’Informaci\'o Cient\'ifica (PIC) data center. PIC is maintained through a collaboration of the Institut de F\'isica d’Altes Energies (IFAE) and the Centro de Investigaciones Energ\'eticas, Medioambientales y Tecnol\'ogicas (Ciemat). We particularly thank Christian Neissner and Martin Børstad Eriksen for their support at PIC. C.P.A., M.R. and N.R. are supported by the ERC via the Consolidator grant ``MAGNESIA'' (No. 817661), the ERC Proof of Concept "DeepSpacePULSE" (No. 101189496), and by the program Unidad de Excelencia Mar\'ia de Maeztu CEX2020-001058-M. V.G. is supported by a UKRI Future Leaders Fellowship (grant number MR/Y018257/1). We also acknowledge support from the Catalan grant SGR2021-01269 (PI: Graber/Rea) and the Spanish grant ID2023-153099NA-I00 (PI: Coti Zelati). C.P.A.’s work has been carried out within the framework of the doctoral program in Physics at the Universitat Autonoma de Barcelona.
\end{acknowledgements}

\bibliographystyle{aa} 

\bibliography{bibliography}
\begin{appendix}
\section{Test experiments}
\label{sec:appendix}
\begin{figure*}
\centering
\includegraphics[width = 0.85\textwidth]{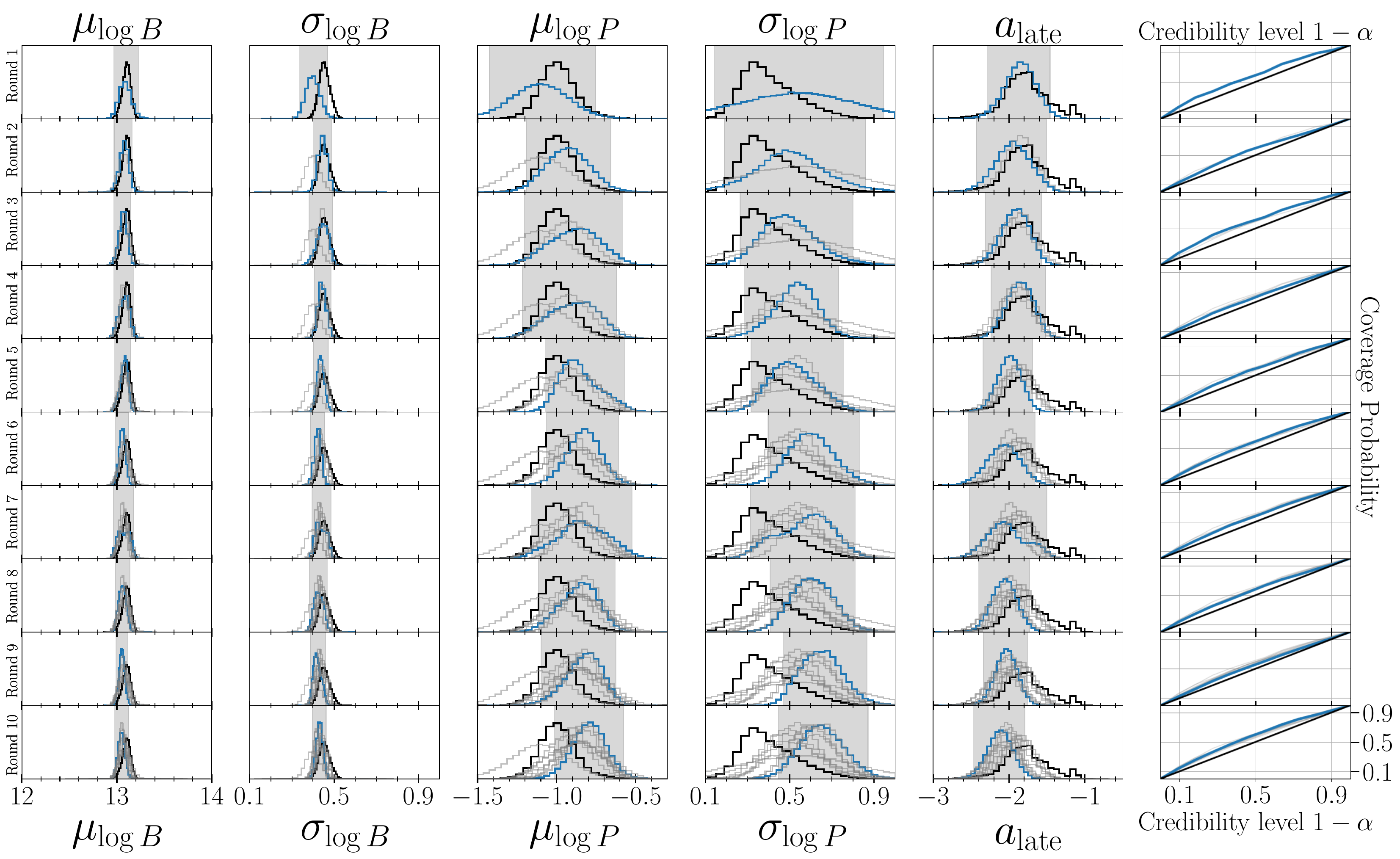}
\caption{Illustration of the \ac{TSNPE} algorithm applied to our pulsar population synthesis. Here, we show results for inferring the five magneto-rotational parameters (as for \citetalias{Graber2024}) of the observed neutron star population in Experiment 1 with 1,000 simulations in round 1. Each row corresponds to one round of inference. The last column shows the coverage probability computed on the test dataset. In each panel, the current round's computed values are shown in blue, while values from previous rounds are shown in light grey. The grey shaded area in the one-dimensional marginal posterior represents the $95\%$ credibility interval of the approximated posterior for that round. In each of these 1D marginal posterior distributions, the horizontal axes represent the parameters’ prior ranges. For comparison, the posterior distribution estimated in \citetalias{Graber2024} is shown as black solid lines. }
\label{fig:exp_1}
\end{figure*}

\begin{figure*}
\centering
\includegraphics[width = 0.85\textwidth]{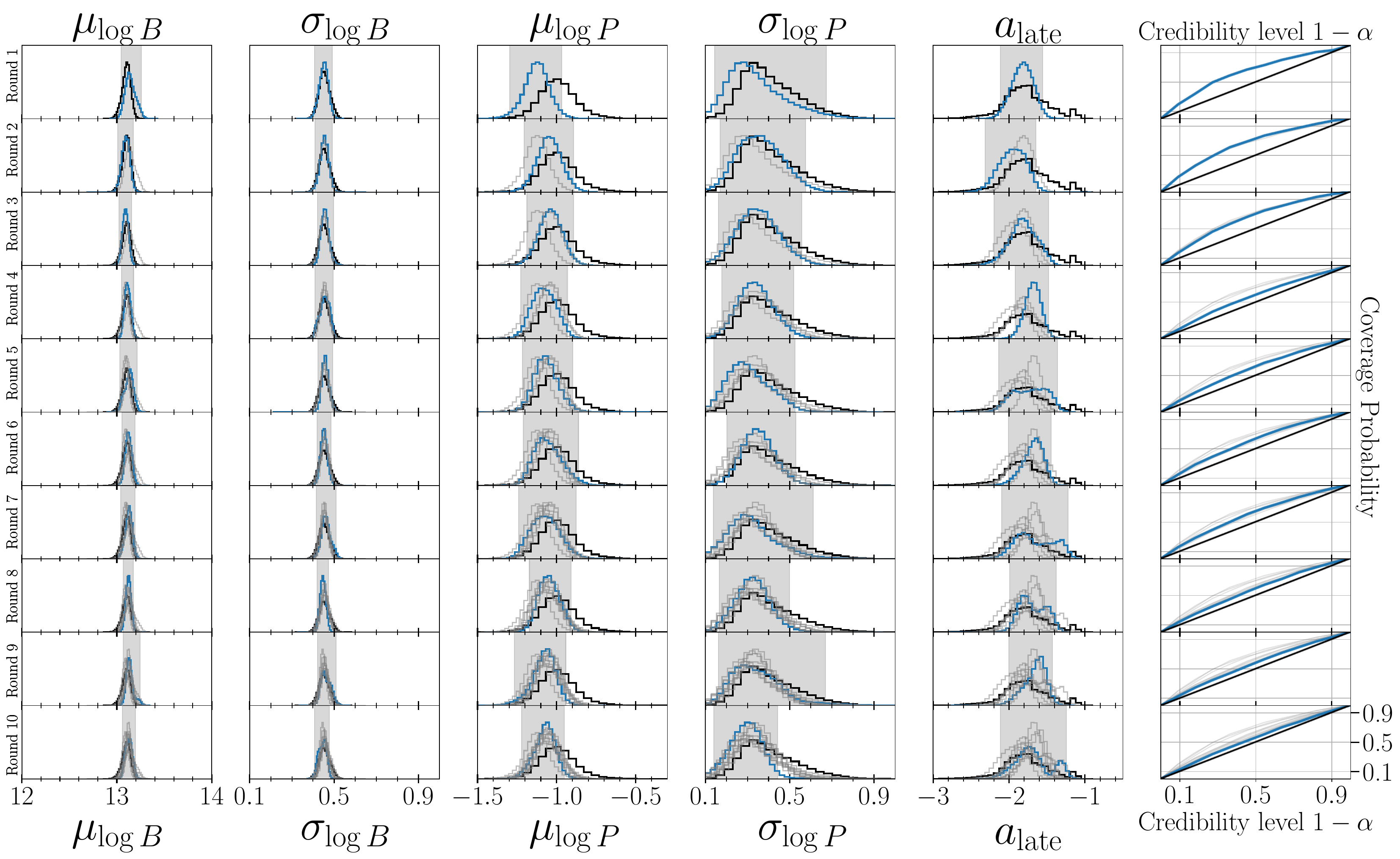}
\caption{Same as Figure \ref{fig:exp_1} but for Experiment 2, where we increased the size of the initial training dataset to $10,000$ simulations.}
\label{fig:exp_2}
\end{figure*}

\begin{figure*}
\centering
\includegraphics[width = 0.9\textwidth]{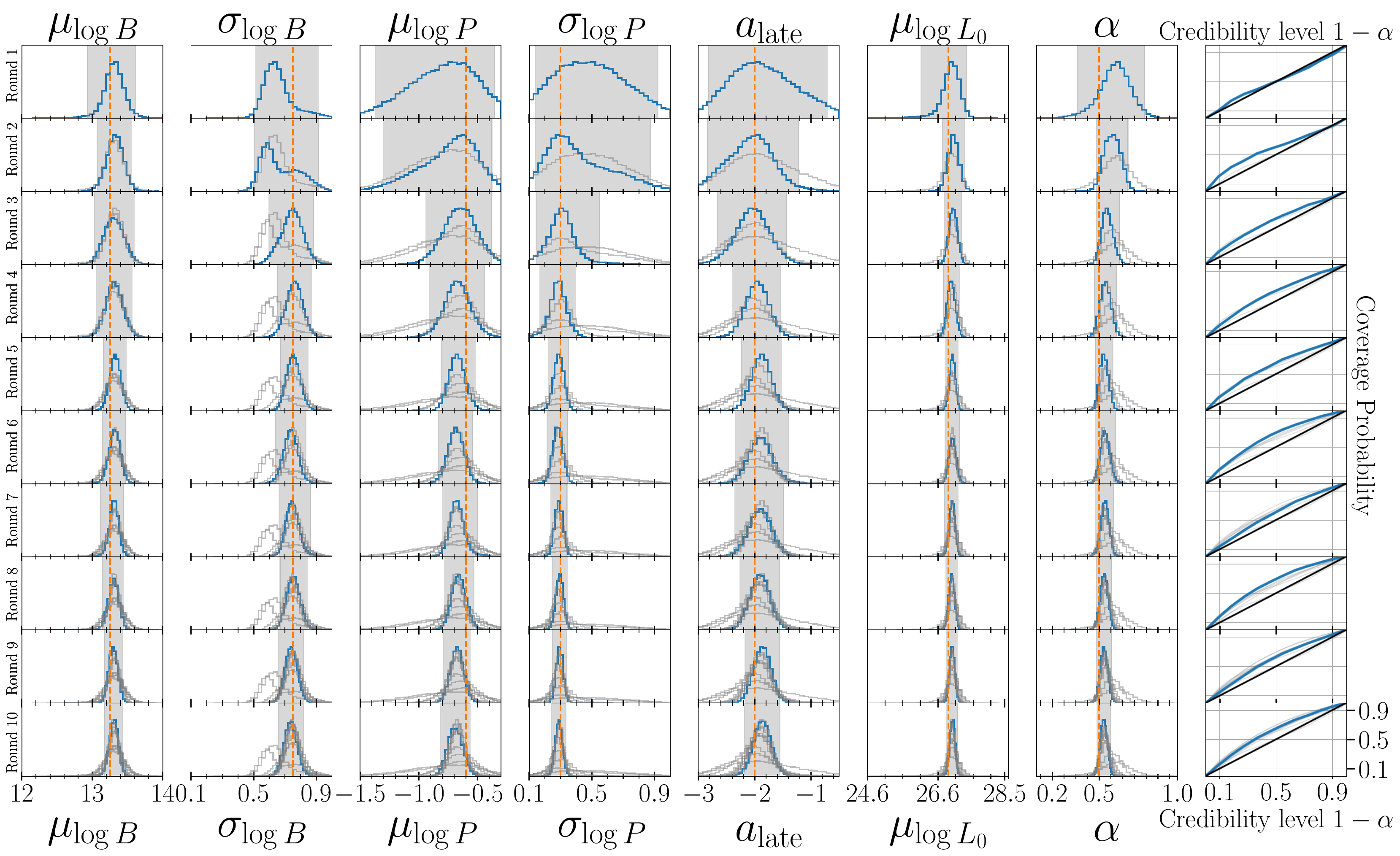}
\caption{Similar to Figure \ref{fig:exp_1} but for Experiment 3, where we applied the \ac{TSNPE} algorithm to infer both the magneto-rotational parameters and the parameters related to the intrinsic bolometric luminosity from Equation \eqref{eqn:luminosity}. This inference was performed for a simulated neutron star population, and the ground truth values, $\bt$, are represented by the orange dashed lines.}
\label{fig:exp_3}
\end{figure*}

Figures \ref{fig:exp_1} to \ref{fig:exp_3} display the results of the TSNPE algorithm over 10 rounds for Experiments 1 to 3, discussed in Sections \ref{subsec:test-5} and \ref{subsec:test-7}. These experiments focus on testing the \ac{TSNPE} methodology with our pulsar population synthesis. In Experiments 1 and 2, we focus on inferring the five magneto-rotational parameters and compare the results with those obtained in \citetalias{Graber2024}. In Experiment 3, we infer the five magneto-rotational parameters along with those related to the intrinsic radio luminosity distribution for a simulated neutron star population with known ground truth, $\bt$. 
\end{appendix}
\end{document}